\documentclass[12pt]{article}
\usepackage[margin=0.5in]{geometry}
\usepackage{amsmath,amssymb}
\usepackage{graphicx}
\usepackage{microtype}
\usepackage{lmodern}
\usepackage[T1]{fontenc}
\usepackage{authblk}
\usepackage{abstract}
\usepackage{titlesec}
\usepackage{hyperref}
\usepackage{orcidlink}  
\usepackage{cite}
\hypersetup{
  colorlinks=true,
  linkcolor=black,
  citecolor=black,
  urlcolor=blue
}

\usepackage{pifont}

\titleformat{\section}{\large\bfseries\sffamily}{\thesection}{1em}{}
\titleformat{\subsection}{\normalsize\bfseries\sffamily}{\thesubsection}{1em}{}
\titleformat{\subsubsection}{\normalsize\itshape\sffamily}{\thesubsubsection}{1em}{}

\newcommand{\E}{\mathbb{E}}
\newcommand{\PP}{\mathbb{P}}
\newcommand{\dd}{\,\mathrm{d}}

\newcommand{\figS}[1]{fig.~S#1}
\newcommand{\figsS}[1]{figs.~S#1}

\title{\sffamily\LARGE\bfseries Social Discounting Enables \\ Fast and Reliable Collective Escape}

\author[1]{Zachary P.~Kilpatrick\,\orcidlink{0000-0002-2835-9416}}
\affil[1]{Department of Applied Mathematics,
  University of Colorado Boulder, Boulder, Colorado 80309, USA (\texttt{zpkilpat@colorado.edu}).}

\date{\today}

\begin{document}
\maketitle

\begin{abstract}
\noindent
Solitary animals face a tradeoff when detecting threats: faster detection means accepting more false alarms. We show that groups can manage this tradeoff better by treating an undisturbed neighbor as evidence against a threat, becoming both faster and more accurate than lone individuals. Modeling each animal as a noisy evidence-accumulator that flees when its belief crosses a threshold, we find that a neighbor's flight signals danger while its stillness signals safety. A naive responder reacts only to flights and inflates false alarms as the group grows; a Bayesian responder weighs both, approximated by a single social discounting rate that interpolates between these limits. This yields closed-form expressions for group performance, including cascade branching ratios that stay strongly subcritical in safety and turn supercritical under threat, so the rate at which an animal discounts a threat while its neighbors stay still can be inferred from behavior alone, and it sets a ceiling on how many neighbors an animal can attend before discounting alone can no longer hold its false-alarm rate. Wild sulphur molly shoals under bird attack are best described by discounting rates well above what individually Bayesian updating supplies over any neighborhood they could plausibly attend, and the same model, at the inferred value, predicts a false-alarm rate that stays constant as shoals grow.
\end{abstract}

When a predator strikes, the difference between escape and capture often depends not on individual reaction time alone but on the collective response of the group. Empirical studies of schooling fish~\cite{herbert2015initiation,rosenthal2015revealing,sosna2019individual}, bird flocks~\cite{papadopoulou2022selforganization,sankey2021absence}, and surface-breathing fish under avian predation~\cite{doran2022fish} have established that escape behavior can spread through animal groups via social amplification. One individual's response influences its neighbors, whose responses influence others, producing rapidly propagating escape waves that can traverse entire groups within seconds~\cite{herbert2015initiation,attanasi2014information}. Such collective displays can involve many thousands of individuals~\cite{doran2022fish}. These waves provide measurable antipredator benefits. Experimentally induced fish waves doubled the interval between successive bird attacks, while more intense waving was associated with reduced capture success~\cite{doran2022fish}; the attack-delaying effect generalizes across avian predators with diverse hunting strategies~\cite{bierbach2025generic}. Collective escape is thus an organized social response with a demonstrable defensive function.

An escape response is a decision under uncertainty since an animal must act before it can know whether a perceived disturbance is truly dangerous. Solitary decision-makers therefore face a speed-accuracy tradeoff, since detecting a threat faster requires acting on weaker evidence and thus admitting more false alarms~\cite{chittka2009speed,bogacz2006physics}. A false alarm wastes energy and interrupts foraging~\cite{lima1990behavioral,lima1999predation}, whereas missed or delayed detection risks predation~\cite{ydenberg1986fleeing,cooper2015escaping}. A recent field study of wild sulphur molly shoals found that larger groups manage this tradeoff better than individuals, becoming both faster and more accurate at distinguishing genuine attacks from harmless disturbances~\cite{ward2011fast,pacher2025better}, a decoupling of true and false positives anticipated by earlier theory~\cite{wolf2013accurate}. We show that a neighbor who has not fled provides evidence against a threat, so the group accumulates social information even when no individual acts.

Collective escape requires individuals to integrate direct environmental evidence with information obtained from their neighbors. Escape waves are initiated by a small number of highly responsive individuals and spread by social influence, with initiation and spreading governed by distinct processes~\cite{herbert2015initiation}, and the interaction networks underlying collective flight are sparse, speed-dependent, and concentrated among nearby neighbors~\cite{rosenthal2015revealing, strandburg2013visual}. Individual escape decisions themselves show signatures of evidence accumulation. Reaction times to looming stimuli follow distributions consistent with drift-diffusion dynamics~\cite{yilmaz2013rapid, evans2018synaptic}, and departure thresholds vary systematically with predation risk and energetic state~\cite{ydenberg1986fleeing, lima1990behavioral}, as predicted by optimal stopping theory. The central challenge is therefore to determine how temporally accumulating private evidence should be combined with the observed actions and inaction of nearby individuals.

A departing neighbor provides positive evidence that a threat is real, whereas continued inaction makes it progressively less likely. Existing models of escape spreading capture only the first effect. They represent collective flight as a social cascade in which influence arrives only with a neighbor's action, so stillness contributes nothing rather than negative evidence, and no agent accumulates private evidence about the threat itself~\cite{rosenthal2015revealing,poel2022subcritical,sosna2019individual,dodds2005generalized}. False alarms are consequently spontaneous cascades, and where their costs are weighed against those of missed detections~\cite{poel2022subcritical,gray2023false}, the weighing is external to the spreading dynamics rather than something the agents perform. Branching ratios and cascade statistics then reflect network structure and a single responsiveness parameter, with no dependence on the state of the world, leaving open how private evidence accumulation shapes collective performance and whether the observed subcritical regime~\cite{poel2022subcritical,gomeznava2023excitable} can emerge from individually rational decisions.

The drift-diffusion model describes two-alternative decisions as noisy evidence accumulation to a threshold that balances speed against error~\cite{wald1947sequential,bogacz2006physics,gold2007neural,ratcliff2016diffusion}. Escape can likewise be viewed as a decision between threat and no threat, consistent with evidence that integrated threat signals trigger escape upon reaching a neural threshold~\cite{evans2018synaptic,evans2019cognitive}. Although drift-diffusion models have been applied to collective choice~\cite{karamched2020bayesian,karamched2020heterogeneity} and social foraging~\cite{bidari2022stochastic}, collective escape has a distinct information structure. A neighbor's departure provides an instantaneous social signal, whereas continued inaction provides evidence gradually.

Here we treat collective threat detection as probabilistic inference, with each animal estimating threat from noisy evidence rather than acting as a passive element of a social cascade~\cite{perezescudero2011collective,sayin2025locusts}. Each agent accumulates private log-likelihood evidence in a drift-diffusion process and departs upon reaching a decision threshold. A naive responder treats a neighbor's departure as a fixed belief jump and ignores the negative evidence conveyed by continued silence, causing false alarms to cascade as group size grows. A Bayesian responder incorporates both, drifting downward during silence and applying a departure jump that increases with delay. We show that this correction is well approximated by a single \emph{social discounting rate} $\alpha$, which interpolates between the naive model ($\alpha=0$) and the large-group Bayesian limit ($\alpha=1$) while remaining analytically tractable. Social discounting improves the collective speed-accuracy frontier, and the threshold required to maintain a fixed group false alarm rate scales as $\frac{\log K}{1+\alpha}$ with the number of attended neighbors $K$. Because $\alpha$ cannot exceed one without driving a genuine detector away from threshold, this scaling caps the number of neighbors an animal can attend at roughly the reciprocal of its false alarm rate before the threshold itself must rise. The same relation yields the empirically observed subcritical regime~\cite{poel2022subcritical} from individual evidence accumulation rather than an imposed constraint.

Because $\alpha$ describes a latent property of individual cognition, it cannot be observed directly in a wild group. We show that it can nonetheless be recovered in closed form from two quantities routinely measured in field studies of collective escape, the group true-positive and false-alarm response rates, in a ratio from which the decision threshold and the noise scale cancel exactly. Applying this inference to sulphur molly data~\cite{pacher2025better} yields $\hat\alpha = 0.95$, above the rate individually Bayesian updating supplies over any neighborhood these fish could attend. The identification requires one further quantity, the number of responders the observed group response pools over, and that count is read from the shape of the latency distribution rather than assumed. The same model, at the fitted value, predicts the constancy of their false alarm rate across shoal sizes. The faster and more accurate collective decisions observed in the wild can thus emerge from private evidence accumulation combined with a strong discounting of the evidence carried by neighbors who have not yet fled.

\begin{figure}[t!]
  \centering
  \includegraphics[width=\linewidth]{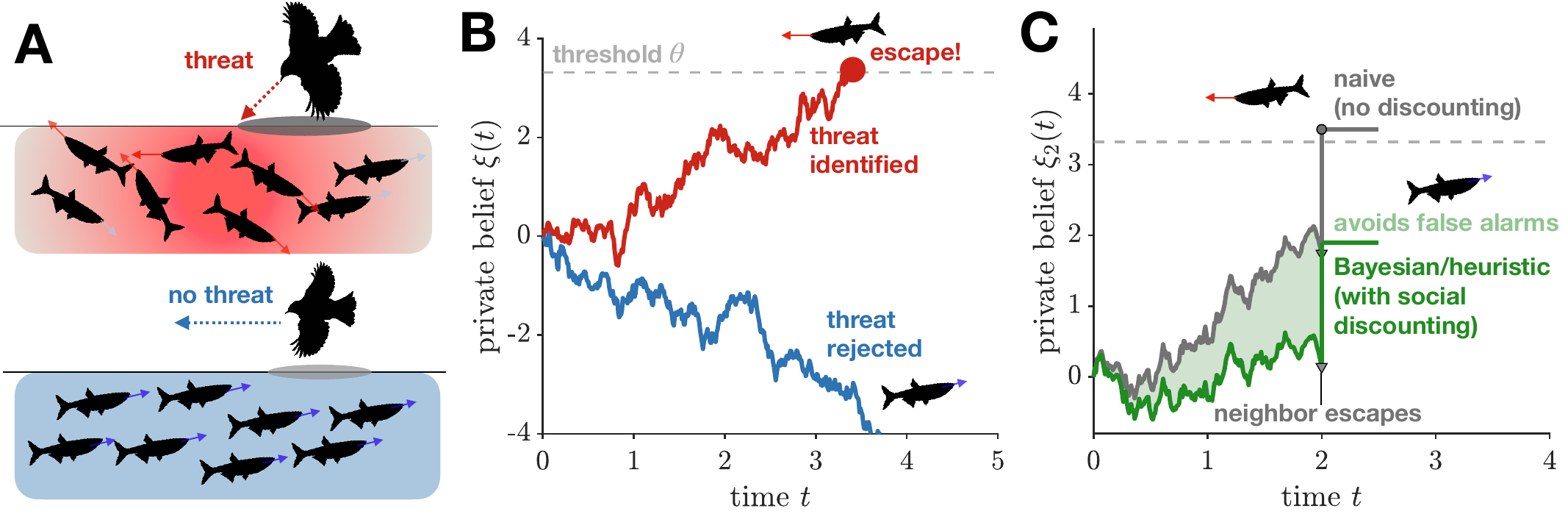}
\caption{\textbf{Combining private and social evidence in collective escape.} Individuals infer whether a threat is present rather than simply copying fleeing neighbors. (\textbf{A}) The same overhead cue, such as a bird's shadow, may signal a real attack (top) or a harmless flyby (bottom). Individuals must therefore accumulate evidence, leading the shoal to scatter under threat but remain calm otherwise. (\textbf{B}) Each individual accumulates noisy evidence until it either reaches a decision threshold and escapes (red) or drifts away and remains (blue). (\textbf{C}) Two responders receive the same belief jump when a neighbor flees but interpret the preceding delay differently. The naive responder (grey) ignores the delay and crosses threshold, producing a false alarm. The discounting responder (green) treats the delay as evidence against threat, so the same jump remains subthreshold. The social discounting rate $\alpha$ interpolates between these rules, with $\alpha=0$ naive and $\alpha=0.8$ shown. Red denotes threat or detection, blue safety, and green social discounting. Panels are illustrative and use a reduced threshold for legibility rather than the fitted $\theta_1$.}
  \label{fig:model}
\end{figure}

\section*{Results}
\label{sec:results}

\subsection*{A model of collective threat detection}
\label{sec:modeloverview}

We model a group of $N$ animals monitoring a shared environment that is either threatening ($H=1$) or safe ($H=0$). Each animal accumulates private sensory evidence as a noisy log-likelihood ratio and flees when it first crosses a fixed threshold $\theta$. Escapes from threat are typically rapid; in safety, they occur only through rare fluctuations that generate false alarms (Fig.~\ref{fig:model}A,B). Animals also observe their neighbors, whose behavior conveys information in two ways. A departure provides immediate evidence of threat, whereas continued stillness provides accumulating evidence of safety. We compare two rules for using this social information. A \emph{naive} responder reacts only to departures, treating each as a fixed jump in belief and ignoring the preceding silence. A \emph{Bayesian} responder uses both, discounting its belief while a neighbor remains and jumping when that neighbor departs (Fig.~\ref{fig:model}C). We show that this Bayesian rule is captured at every group size by a single \emph{social discounting rate} $\alpha\in[0,1]$, with $\alpha=0$ for the naive rule and $\alpha=1$ in the large-group Bayesian limit. This parameter, which can be inferred from field data, determines how collective speed and accuracy scale with group size. Full derivations and numerical methods appear in Materials and Methods.

\subsection*{The single-agent speed-accuracy frontier}
\label{sec:frontier_single}
\begin{figure}[t!]
  \centering
  \includegraphics[width=\linewidth]{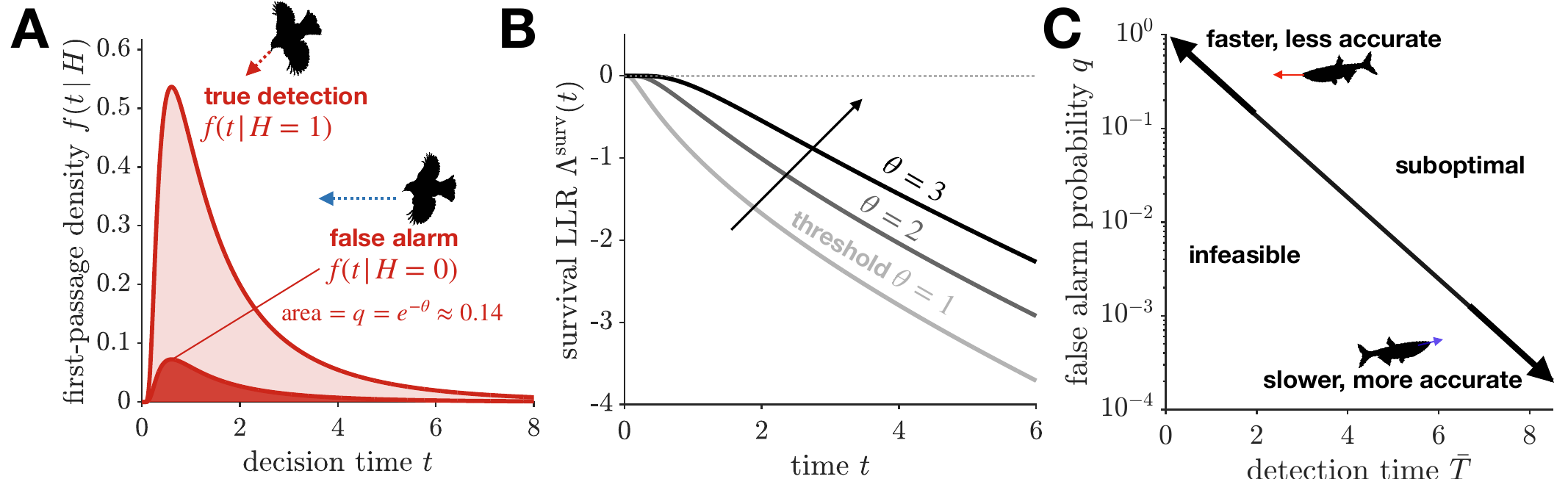}
\caption{{\bf Evidence accumulation and the speed--accuracy tradeoff for a single individual.} Model parameters $\mu_1 = |\mu_0| = 1$, $\sigma = \sqrt{2}$, at an illustrative $\theta = 2$. (\textbf{A})~The time an individual takes to decide follows a skewed distribution. Under threat (light red, a striking bird) it always eventually departs; with no threat (dark red, a harmless flyby) it departs only through a rare fluctuation, with probability $q = e^{-\theta}$ at the threshold shown. Both distributions are skewed toward later times, so the rare false alarms tend to come especially late. (\textbf{B})~The longer a neighbor stays in place, the stronger the evidence that no threat is present. That accumulating evidence, measured by the survival log-likelihood ratio, grows more steeply when individuals require more evidence to act ($\theta = 1, 2, 3$, light to dark), and it is what the social model below builds on. (\textbf{C})~A lone individual cannot detect threats both faster and more reliably. Lowering the threshold speeds detection but raises the false alarm rate (upper left), and raising it does the reverse (lower right). The curve is the best available tradeoff, with combinations below it unreachable and those above wasteful. A group using social information pushes it inward (Fig.~\ref{fig:largeN}).}
  \label{fig:singleagent}
\end{figure}

A lone agent sets the baseline against which group performance is measured. Because $\xi$ is a log-likelihood ratio, an agent with threshold $\theta$ escapes a real threat after a mean time $\bar T=\theta$ and exhibits a false alarm in the absence of a threat with probability $q=e^{-\theta}$ (Fig.~\ref{fig:singleagent}A; Materials and Methods). The same two distributions determine what an agent's behavior reveals to a neighbor, since inaction favors the safe hypothesis, providing an accumulation of negative evidence measured by the survival log-likelihood ratio (Fig.~\ref{fig:singleagent}B). For the lone agent itself only the two rates matter. It has no neighbors to discount, so its false alarm rate carries no dependence on $\alpha$, and it is the value a group must hold fixed as it grows, fixing the threshold scaling derived below. Every threshold therefore yields the operating point $\bigl(\bar T(\theta),q(\theta)\bigr)=\bigl(\theta,e^{-\theta}\bigr)$, and as $\theta$ ranges over $(0,\infty)$ these points sweep out a convex frontier (Fig.~\ref{fig:singleagent}C). Faster detection carries an exponential cost in accuracy, with $q=e^{-\bar T}$, providing a straightforward speed-accuracy tradeoff. The operating point depends on how an agent weighs missed or delayed detection against false alarms. We do not posit this weighting because it is not separately identifiable from field data, and the empirical analysis below works directly with the observable coordinates $\bar T$ and $q$. What matters is the achievable set and how social coupling deforms it. Our central finding is that it shifts this curve strictly inward, allowing a group to become simultaneously faster and more accurate than a lone agent; social discounting then determines how much threshold, and hence how much speed, a group of a given size must spend to hold its accuracy there.

\subsection*{Escape cascades in dyads}
\label{sec:dyad}

\begin{figure}[t!]
  \centering
  \includegraphics[width=\textwidth]{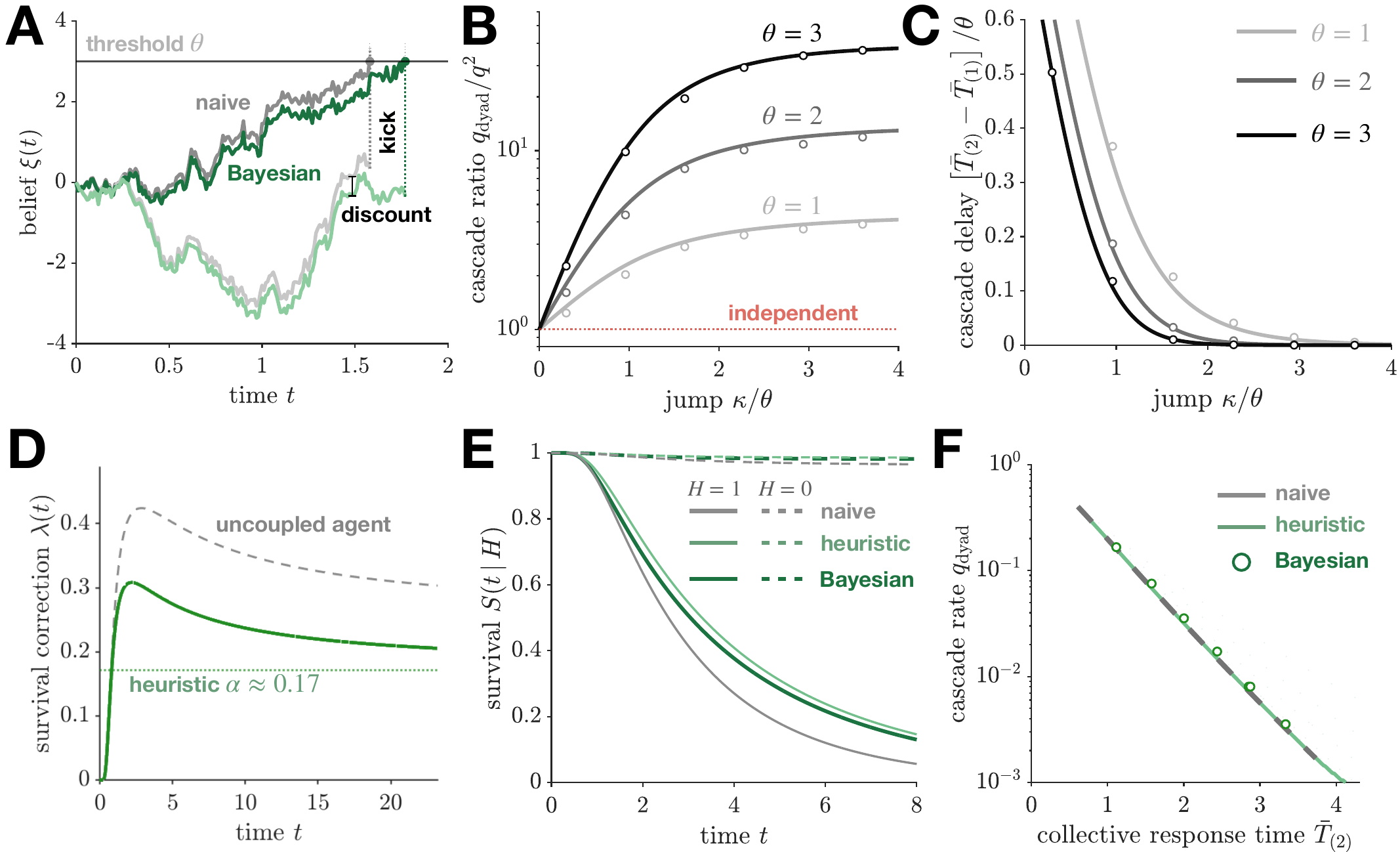}
  \caption{\textbf{The cost of ignoring a neighbor's hesitation, and how discounting it restores accuracy.} Two individuals accumulate private evidence and flee at a threshold; when one departs, the other's belief jumps by a fixed $\kappa$ under the naive rule, while the Bayesian rule reads the delay as evidence against a threat. Naive is grey and Bayesian green wherever the two are compared (A, E, F); in B and C the grey shades run light to dark over $\theta = 1,2,3$. Throughout $\mu_1 = |\mu_0| = 1$ and $\sigma = \sqrt2$, with $\theta_1 = 3.568$ in D and E, an illustrative $\theta = 3$ in the schematic A, and a sweep over $\theta$ in F. (\textbf{A})~Two pairs of trajectories when no threat is present, each pair on shared noise. Above, the fixed kick carries the naive responder over threshold and turns one false alarm into a group-wide one, while the accumulated discount holds the Bayesian responder below; below, neither departs. (\textbf{B})~Larger jumps inflate the false-alarm cascade probability, shown as the ratio $q_{\rm dyad}/q^2$ against the independent baseline (red dotted). (\textbf{C})~The same jumps shorten the threat-induced cascade delay $[\bar T_{(2)} - \bar T_{(1)}]/\theta$, the speed that error buys. (\textbf{D})~The Bayesian discount rate $\lambda(t) = h_1(t) - h_0(t)$ (green) lies below the uncoupled hazard gap (grey dashed) at all times and, after an early peak, relaxes toward the closed-form ceiling $L_\infty(2) = 3 - 2\sqrt2 \approx 0.17$ (dotted). (\textbf{E})~Survival $S(t \mid H)$ under threat (solid) and no threat (dashed). The constant-$\alpha$ heuristic tracks the exact Bayesian rule to $2.6\times10^{-2}$ and $1.1\times10^{-3}$ respectively, while the naive rule deviates in the threat context. (\textbf{F})~Speed--accuracy comparison, parametric in $\theta$. The three rules lie close together at $N=2$, with the exact Bayesian dyad above the naive curve rather than below it by four to seventeen percent in $q_{\rm dyad}$, three to five Monte Carlo standard errors at the lower thresholds. Lines are analytic in B and C and from the fixed-point sweep in F; circles are Monte Carlo, $n = 10^4$ in B and C and $n = 4\times10^4$ in F with a Broadie--Glasserman--Kou discrete-barrier correction~\cite{broadie1997continuity}.}
  \label{fig:dyad}
\end{figure}

Pulsatile social coupling, in which a neighbor's decision produces a fixed belief update, has been studied in collective two-alternative choice~\cite{caginalp2017decision}, quorum models~\cite{reina2023asynchrony}, social patch foraging~\cite{bidari2022stochastic,blummoyse2025egalitarian,blummoyse2026hierarchy}, and escape-wave spreading~\cite{rosenthal2015revealing,poel2022subcritical}. Rational agents on a network can instead extract information from a neighbor's continued silence, which grows informative over time when thresholds are asymmetric~\cite{karamched2020bayesian}, or other neighbors have decided already~\cite{karamched2020heterogeneity}. Those settings are irreversible two-alternative choices in which every agent is deciding between the same pair of options. Escape differs in that social signals push only toward action and the two errors are not symmetric, a false alarm wasting energy while a missed detection risks predation, and it is that asymmetry that makes silence evidence for one hypothesis specifically.

An agent using the naive pulsatile model responds to the other's departure with a fixed belief jump $\kappa$, so one agent's false alarm can carry its neighbor over threshold (Fig.~\ref{fig:dyad}A). This produces an inherent tradeoff. Increasing $\kappa$ shortens the cascade delay when $H=1$, so social signals accelerate the group response, but inflates the false-alarm cascade probability when $H=0$ from the independent baseline $q^2$ toward the single-agent rate $q$ (Fig.~\ref{fig:dyad}B,C). Neither extreme is desirable, since $\kappa=0$ gives independent agents with no social benefit while large $\kappa$ converts every individual false alarm into a group one. The amplification arises because the naive rule ignores the information in a neighbor's continued non-departure, which is negative evidence against the threat (Fig.~\ref{fig:singleagent}B). The Bayesian rule instead discounts each agent's belief at rate $\lambda(t)=h_1(t)-h_0(t)$ during the other's silence, both agents reading the other's stillness as evidence of safety.

The information carried by this silence depends on how likely the neighbor was to have departed by then, which itself depends on the discounting applied to the neighbor, so the correction must be determined self-consistently. The dyad therefore satisfies the fixed-point equation $\lambda=h_1[\lambda]-h_0[\lambda]$, which we solve numerically (Materials and Methods). Two features of the solution matter. First, after an early transient $\lambda(t)$ relaxes toward a constant (Fig.~\ref{fig:dyad}D), supporting the closed-form approximation developed below. Second, that constant lies well below the uncoupled single-agent hazard gap: mutual discounting suppresses both hazards and narrows their difference, making the coupling self-limiting rather than runaway.

Where a specific value is needed we take the late-time limit $L_\infty(2) = 3-2\sqrt2 \approx 0.17$ of Eq.~(\ref{eq:Linf}), the model's own value in closed form, independent of the threshold and the integration horizon, and the same rule we apply at every group size. Little rests on that choice, since $\alpha$ is a free parameter of the heuristic and is fitted to data below; the convention matters only where we compare the heuristic against the exact Bayesian dyad at matched $\alpha$. There the limit is not the most accurate constant available, because the decision plays out while $\lambda(t)$ is still above it, and the limit and a window mean bracket the exact first-departure time from either side (Materials and Methods).

At any fixed threshold, the Bayesian dyad is more accurate. The anti-drift $-\lambda(t)$ pushes each agent's false alarm probability below the uncoupled value $e^{-\theta}$ and cuts the cascade probability $q_{\rm dyad}$ by about two fifths at the threshold we fit to the molly data below. Both rules are visible in the survival functions, where the naive rule departs sooner under threat while the constant-$\alpha$ heuristic tracks the exact Bayesian rule closely (Fig.~\ref{fig:dyad}E). The two rules barely separate on the speed--accuracy frontier itself, however (Fig.~\ref{fig:dyad}F). The collective response time is the expected \emph{maximum} of two coupled first-passage times, so the same discounting that suppresses false alarms also delays the first departure when $H{=}1$, and for the dyad the two effects very nearly cancel.

What separation there is runs against the Bayesian rule, because the two rules optimize different objectives. Bayesian updating minimizes an \emph{individual} agent's inference loss, whereas the frontier pairs cascade probability with the collective response time $\bar T_{(2)}=\E[\max(T_1,T_2)\mid H{=}1]$, which depends on the joint first-passage structure of the pair. Decoupling $\alpha$ from exact belief updating lets the heuristic be tuned to that objective directly, and for the dyad this buys almost nothing; the exact Bayesian rule does not lie on the tuned envelope at all, sitting above the naive curve at matched response time by four to seventeen percent in $q_{\rm dyad}$, three to five Monte Carlo standard errors at the lower thresholds. The effect is small, but its sign is the point, and it follows from the mismatch of objectives rather than any deficiency of the inference. The dyad is thus the regime in which the naive rule is effectively optimal, and the separation opens only as $N$ grows.

\subsection*{Collective escape in $N$-agent groups}
\label{sec:largeN}

The dyad analysis motivates a heuristic $(\alpha,\theta)$ model as a closed-form approximation to the Bayesian fixed point, replacing the time-varying survival correction by a constant downward drift $-\alpha$ during a neighbor's silence and a time-dependent kick on departure (Materials and Methods). The approximation holds at every group size we examine, and larger groups employing this heuristic escape both more accurately and sooner (Fig.~\ref{fig:largeN}C). At matched false alarm rate, discounting lowers the threshold a group requires, so it detects a genuine threat faster.

\subsubsection*{The heuristic is the large-$N$ Bayesian limit}

\begin{figure}[t!]
  \centering
  \includegraphics[width=0.75\linewidth]{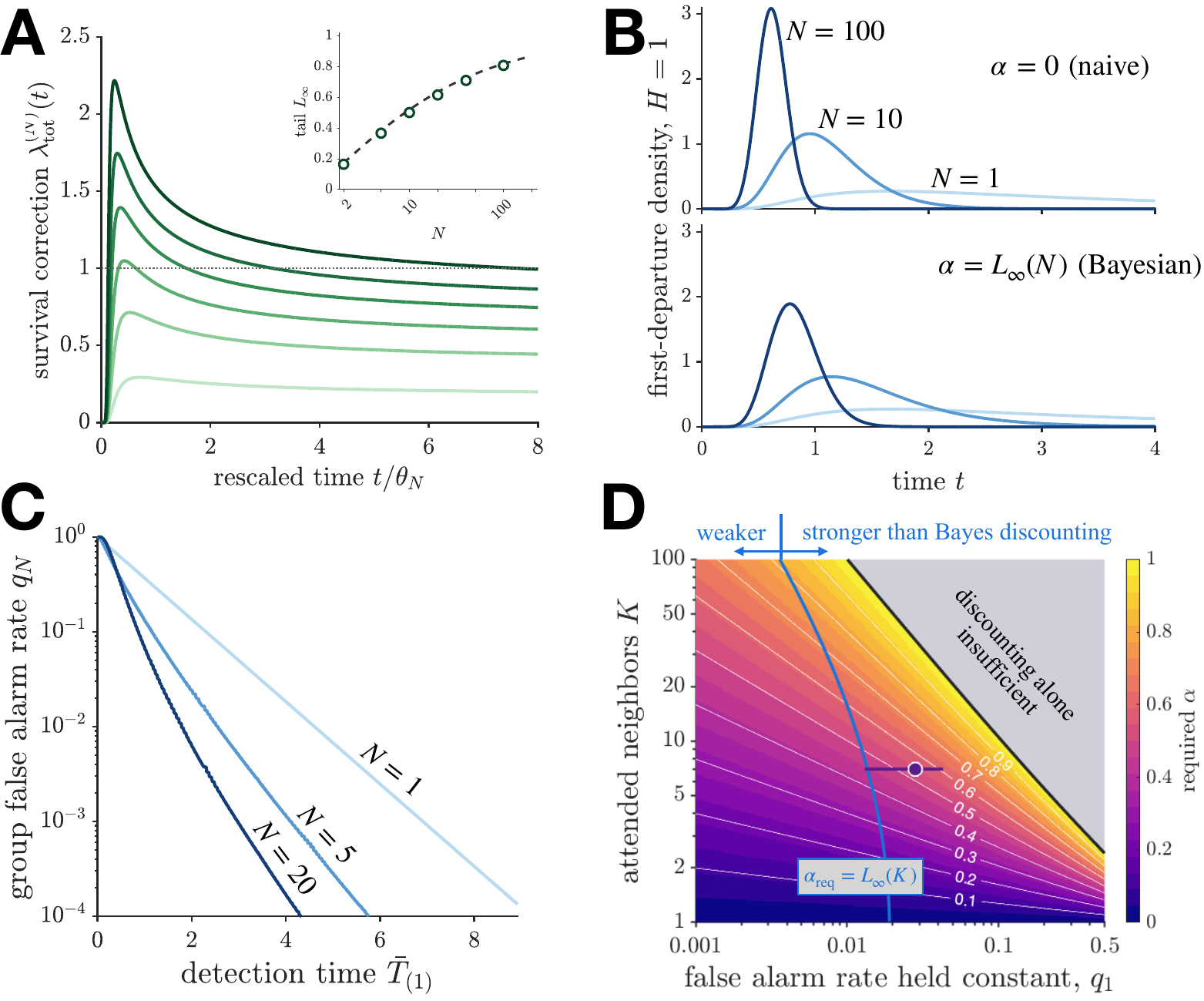}
  \caption{\textbf{Collective escape in $N$-agent groups using the heuristic $(\alpha,\theta)$ model.} $\mu_1 = |\mu_0| = 1$, $\sigma^2 = 2$, $\theta_1 = -\log q_1 = 3.568$ from the per-unit false alarm rate $q_1 = 0.028$. Times are in model units. (\textbf{A})~Total survival drift $\lambda^{(N)}_{\rm tot}(t) = (N{-}1)\lambda^{(N)}(t)$ against rescaled time for $N=2,5,10,20,40,100$ (light to dark green), all six solved at the single threshold $\theta_1$. Each curve decays toward its late-time limit from above. Inset, the extrapolated tails (markers) against the closed form of Eq.~(\ref{eq:Linf}) (dashed), which they approach from below as $N$ grows; dotted, the $N\to\infty$ limit. The early-time peak is a transient that instead diverges as $N/(\log N)^2$ (\figS{1}). (\textbf{B})~Group first-departure density $f_{(1)}(t\mid H{=}1)$ at $\theta_1$ for $N=1,10,100$ (light to dark blue), under the naive rule $\alpha=0$ (top) and the self-consistent rate $\alpha = L_\infty(N)$ (bottom); $L_\infty(1)=0$, so $N=1$ is common to both. Mean detection time falls from $3.57$ to $0.62$ under the naive rule. Both cases fix the threshold, so discounting increases accuracy but reduces speed; Eq.~(\ref{eq:threshold_scaling}) fixes the accuracy instead, and there discounting buys speed. (\textbf{C})~Group Pareto frontier $(\bar{T}_{(1)}, q_N)$ for $N=1,5,20$, each the lower envelope over $\alpha$ and $\theta$. The frontiers are strictly nested, and at matched $q_N=10^{-2}$ a group of $20$ detects $2.6\times$ faster than a lone agent. (\textbf{D})~Discounting rate required to hold the group false alarm rate at $q_1$ as the attended count $K$ grows, at fixed threshold; white contours at $0.1$ intervals. Grey, where the requirement exceeds one, so the threshold must rise as well. Blue, where it equals the individually Bayesian rate $L_\infty(K)$, a critical false alarm rate of order one percent nearly independent of $K$; to its right, holding the observed rate demands more discounting than Bayesian updating supplies. The sulphur molly operating point $(q_1, K) = (0.028, 7)$ lies there, at $\alpha_{\rm req} = 0.54$ against $L_\infty(7) = 0.45$, which the empirical fit reaches independently (Fig.~\ref{fig:branching}); the horizontal bar identifies the interval in $q_1$ implied by that on $\hat M$ of Eq.~(\ref{eq:Mfit}) under the pooling inversion of Materials and Methods.}
  \label{fig:largeN}
\end{figure}

Silence in a large group is not much more informative than silence in a small one. Each additional quiet neighbor supplies less evidence than the last, so the reassurance a Bayesian agent accumulates levels off rather than growing without bound. In the model, the total survival correction $(N-1)\lambda^{(N)}(t)$ saturates in group size (Fig.~\ref{fig:largeN}A): after an early transient it relaxes monotonically toward a finite late-time value, and that value rises with $N$ toward a hard ceiling. Replacing $\lambda^{(N)}(t)$ by a fixed effective drift, the defining move of the heuristic, therefore introduces no qualitative error. The early-time peak, by contrast, is a transient that diverges as $N/(\log N)^2$ (Supplementary Materials).

The sustained value of this correction saturates at unity, and the reason is behavioral. An agent's own evidence for a threat arrives at unit rate; the discount from its neighbors' silence pushes the other way. A discount held above unity would drive a genuine detector steadily away from threshold, so a group in real danger would talk itself out of fleeing. The early transient does briefly exceed one, but it is short-lived and the displacement it produces is bounded; what cannot persist is a late-time value above the balance point. At that balance a true-positive straggler that has not yet departed is held at the boundary, its private evidence cancelled by the accumulated weight of its still-silent neighbors.

The saturated value has an exact closed form, determined by a self-consistency condition: how much a silence should discount an agent's belief depends on how fast its neighbors would have departed, which depends in turn on how much their own silences discount theirs. Writing $L^{(N)} \equiv (N-1)\lambda^{(N)}$ for the total survival drift and $L_\infty$ for its late-time value, that condition is a quadratic with physical root
\begin{align}
  L_\infty(N) = \frac{(k+2) - 2\sqrt{k+1}}{k}, \qquad k = N-1,
  \label{eq:Linf}
\end{align}
derived in Materials and Methods and confirmed by the numerical fixed point at every $N$ (Fig.~\ref{fig:largeN}A, inset). At $N=2$ this gives the dyad value $3-2\sqrt2 \approx 0.17$, and $L_\infty(N) = 1 - 2/(\sqrt N + 1)$ exactly, so the ceiling is reached only in the infinite-group limit, and slowly enough that real groups sit well short of it. Identifying $\alpha$ with this saturated drift makes the heuristic a one-parameter interpolation between the naive rule ($\alpha=0$) and the strict large-$N$ Bayesian limit ($\alpha=1$), with the finite-$N$ member fixed in closed form.

Figure~\ref{fig:largeN}B contrasts the naive model with the model's own self-consistent rate $\alpha = L_\infty(N)$ at fixed $\theta=\theta_1$. In both, the first-departure density shifts sharply leftward and concentrates as $N$ grows, the naive mean falling from $3.57$ at $N=1$ to $0.62$ at $N=100$: pooling $N$ independent evidence streams speeds detection as $1/\log N$. Discounting preserves most of that gain at a price in raw speed, about a third at $N=100$, in exchange for the false alarm suppression quantified below. Because this comparison holds the threshold common to both rules, discounting buys accuracy and pays in speed; under the threshold scaling below, accuracy is fixed by construction and discounting buys speed instead.

\subsubsection*{Speed-accuracy Pareto frontier}
\label{sec:pareto}

For a group of $N$ agents the tradeoff is the frontier swept out in $(\bar{T}_{(1)}, q_N)$ space, where $q_N = 1-(1-q_\alpha(\theta))^N$. Sweeping $\theta$ and $\alpha$ at each group size and retaining the non-dominated points gives the frontier (Fig.~\ref{fig:largeN}C). The frontiers for $N=1,5,20$ are strictly nested with no crossover, so larger groups reach a lower false alarm rate at every detection time. Pooling therefore improves collective performance uniformly across the plane, and most where it matters for survival, in the regime of rare false alarms: at $q_N=10^{-2}$ a group of $20$ detects $2.6\times$ faster than a lone agent.

\subsubsection*{Threshold scaling and subcriticality}
\label{sec:threshold_scaling}

A lone agent has no neighbors to discount, so its false alarm rate is $q_1 = e^{-\theta_1}$, with no dependence on $\alpha$, and we take this as the rate a group must hold as it grows. The reason is economic: a spurious dive costs an individual the same interrupted foraging whether it is one fish among ten or one among ten thousand, so the tolerable error rate is set by what an individual can afford, not by how many neighbors it happens to have. Holding it is not automatic, since any one of $N$ agents can trigger a departure and the group rate $q_N \approx N q_\alpha(\theta)$ amplifies $N$-fold at fixed threshold. Pinning it at the solitary value requires
\begin{align}
  \theta_N = \frac{\theta_1 + \log N}{1+\alpha},
  \label{eq:threshold_scaling}
\end{align}
which cancels that amplification to leading order. The threshold grows as $\log N/(1+\alpha)$, a slope interpolating between $1$ for the naive rule and $1/2$ in the large-$N$ Bayesian limit, and at fixed group size falls as $1/(1+\alpha)$, so a strong discounter reaches the same accuracy from a much lower barrier. This demonstrates the advantage of discounting, since with accuracy fixed by construction the benefit appears entirely as speed. At $N=20$ the group detection time falls from $\bar T_{(1)} = 1.77$ under the naive rule to $0.95$ at the rate $\hat\alpha = 0.95$ we fit below, an acceleration of $1.9\times$ at no cost in accuracy. When each agent attends a fixed local neighborhood the relevant count is $K$, giving $\theta_K = (\theta_1 + \log K)/(1+\alpha)$, consistent with the behavior of sulphur mollies as we show below. Their false alarm rate does not increase with shoal size, as it would in a group that held its threshold fixed.

At this threshold the cascade branching ratios simplify. When $H=1$ every surviving neighbor sits near threshold, so the kick clears all of them and $b_{H=1} = K$ in the mean-field limit, where $K=N-1$ in the all-to-all case. When $H=0$ the higher threshold offsets each added neighbor exactly, giving $b_{H=0} = K\,q_\alpha(\theta_K) = q_1$, independent of $K$ and of $\alpha$, with the fixed level the solitary false alarm rate itself. The cascade is thus supercritical under threat and strongly subcritical at every group size, and their ratio $b_{H=1}/b_{H=0} = K/q_1$ grows linearly with the attended count. The collective therefore amplifies real signals and damps spurious ones, with a discriminability set by how rare a solitary false alarm is.

Holding the false alarm rate as $K$ grows requires either a rising threshold or stronger discounting, and Fig.~\ref{fig:largeN}D shows how the two trade off. At fixed threshold the requirement is $\alpha_{\rm req} \approx \log K/\log(1/q_1)$ (Eq.~\ref{eq:alpha_req}), and since a rate above one would drive a genuine detector away from threshold, that requirement bounds the attended count at roughly $1/q_1$, a ceiling real groups sit well inside (Materials and Methods). The tighter constraint is individual inference, since $\alpha_{\rm req}$ exceeds the individually Bayesian rate $L_\infty(K)$ above a critical false alarm rate near one percent, nearly independently of $K$. Any animal noisier than that must discount more steeply than Bayesian updating supplies to attend more neighbors without inflating false alarms. Where a species falls is an empirical question, and the sulphur molly shoals we fit below sit just above that rate, so their false alarm rate alone anticipates the conclusion the response-rate identification reaches independently.

\subsection*{Empirical measurement of social discounting}
\label{sec:branching}

The heuristic model has two free parameters, the social discounting rate $\alpha\in[0,1]$ and the departure threshold $\theta>0$. The central empirical claim is that $\alpha$, a latent property of individual cognition, can be recovered from group-level escape rates alone, without belief trajectories, cost measurements, or controlled manipulation. The identification uses the group true- and false-positive response rates in a ratio from which the threshold and the accumulator noise scale cancel, so it assumes nothing about the signal-to-noise ratio of the underlying evidence accumulation (Materials and Methods). It needs one further quantity, the number of units the group response pools over, and the latency distribution supplies it through a scale-free ratio that rises steeply with the pooling count at a threshold the false alarm rate fixes rather than leaves free. Fitting wild sulphur molly shoals~\cite{pacher2025better} gives a discounting rate above what individually Bayesian updating supplies at any admissible neighborhood, and the same model at that value predicts a false alarm rate constant in shoal size, a scaling law that plays no part in the fit.

\subsubsection*{Identifying $\alpha$ and the pooling count}
\label{sec:identification}

The cleanest identification of $\alpha$ uses the two group-level response probabilities and is independent of both threshold and noise scale. Pooling over the $M$ responders whose earliest crossing registers as a group response recovers the per-unit false alarm and miss probabilities from the group rates (Materials and Methods). In the heuristic model these are the two threshold-crossing probabilities given the shifted drifts $-(1+\alpha)$ and $(1-\alpha)$, and the ratio of their logarithms cancels the common scale factor exactly,
\begin{align}
  r \equiv \frac{\log q_{\rm ind}}{\log \mathrm{miss}_{\rm ind}} = \frac{1+\alpha}{1-\alpha}
  \qquad\Longrightarrow\qquad
  \hat\alpha = \frac{r-1}{r+1}.
  \label{eq:sigmafree}
\end{align}
Neither the threshold $\theta$ nor the noise scale $\sigma^2$ appears, so $\alpha$ is identified from the two response rates alone once the count they pool over is known. That count is not free either. The latency distribution has an inverse-Gaussian shape parameter $\lambda$ and a mean, both carrying units of time, so their ratio $\lambda/\bar T_{(1)}$ is dimensionless. For the minimum over a pool of accumulators this ratio rises steeply with the size of the pool. It also depends on the threshold, but the threshold is not free, since naming $M$ fixes it at $\theta_1(M) = -\log[1-(1-q)^{1/M}]$ through the measured false alarm rate. The calibration therefore runs along a one-parameter family in $M$ alone and the observed ratio determines $M$ (\figS{3A}).

Three measurements of group size enter our analysis, and must be kept separate. The \emph{attended count} $K$ is the number of conspecifics a given fish monitors, set by sensory range rather than by shoal size. It controls the threshold and is the argument of the individually Bayesian rate $L_\infty$, and it is not identified by the group rates. The \emph{pooling count} $M$ is the number of effectively independent responders whose earliest crossing registers as a group response, and it is what the rate inversion inverts over. It is not the shoal census. A repeat wave is recorded when any fish responds, which would make the raw pool the whole shoal, but fish packed at close spacing share correlated and occluded visual fields, so the number of effectively independent detectors is far smaller. Nothing in the identification assumes otherwise, and $M$ is read from the latency shape rather than set to the shoal size; the fitted value is of order $10^{-4}$ of the fish present. The \emph{shoal census} $N$ is that full population, which grows without changing $K$. This asymmetry, local control of false alarms together with pooling of detections over a count that grows with the shoal, is what permits detection to accelerate while the false alarm rate stays flat.

\begin{figure}[t]
\centering
\includegraphics[width=\linewidth]{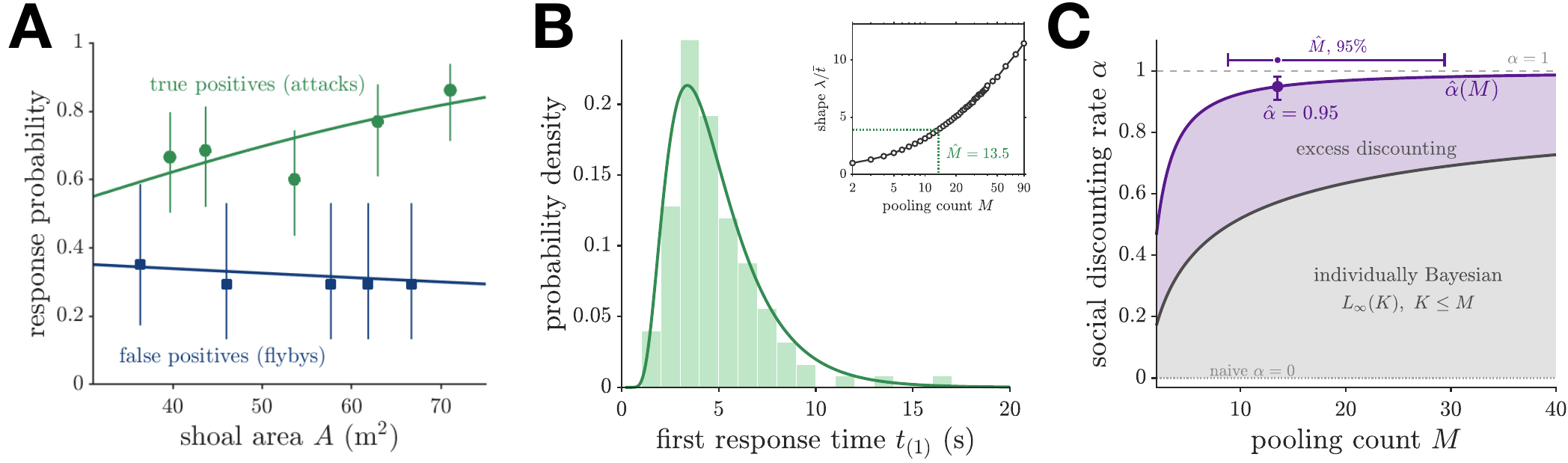}
\caption{\textbf{Social discounting recovered from wild sulphur molly shoals.}
(\textbf{A}) Group response probability against shoal area for kiskadee attacks ($n=177$, green circles) and harmless flybys ($n=81$, blue squares), binned with Wilson intervals and shown with logistic fits. The flyby rate is flat (slope $-0.006$, SE $0.021$, $P=0.78$) while the attack rate rises ($+0.033$, SE $0.014$, $P=0.019$).
(\textbf{B}) Latency to the first repeat wave ($n=125$, $\bar{T}_{(1)}=4.92$~s) with the maximum-likelihood inverse-Gaussian density. Inset, the scale-free ratio $\lambda/\bar{T}_{(1)}$ against pooling count from a Monte Carlo calibration of the minimum of $M$ first passages, run at the self-consistent threshold $\theta_1(M)=-\log[1-(1-q)^{1/M}]$ so that nothing is tuned. The observed value $3.93$ (dotted) gives $\hat{M}=13.5$ ($95\%$ CI $[8.8,\,29.4]$).
(\textbf{C}) Discounting rate against pooling count. Purple, $\hat\alpha(M)$ from Eq.~(\ref{eq:sigmafree}) at the observed response rates, with the point estimate $\hat\alpha=0.95$ and its joint bootstrap interval, resampled over recording-by-bout clusters and stratified by event type. Grey, the individually Bayesian ceiling $L_\infty(K)$ over all admissible attended counts $K\le M$. The shaded band between them is discounting in excess of what Bayesian updating supplies at any neighborhood, $0.38$ at the point estimate and never below $0.20$ across $4000$ replicates. Bar at top, the $95\%$ interval on $\hat{M}$.}
\label{fig:branching}
\end{figure}

\subsubsection*{Fitting to sulphur molly shoals}
\label{sec:pacherfitting}

We fit the model to wild sulphur molly (\textit{Poecilia sulphuraria}) shoals responding to aerial attacks by great kiskadees (\textit{Pitangus sulphuratus})~\cite{pacher2025better}. Kiskadee attacks closely resemble harmless flybys, since the bird only briefly dips its beak to the water surface, so the shoal faces a genuinely ambiguous stimulus. The collective response is a repeat wave, in which fish dive and resurface in a propagating pattern that serves as an antipredator signal. A true positive is a repeat wave following a real attack, a false positive is a repeat wave following a flyby, and a genuine attack drawing no repeat wave within the observation window is a false negative.
Across $n = 177$ kiskadee attacks the group true-positive rate is $\mathrm{TP} = 0.718$, with $127$ responses and $50$ false negatives, and across $n = 81$ flybys the group false-alarm rate is $q = 0.321$, with $26$ responses and an unclustered Wilson $95\%$ interval of $[0.229,\,0.429]$. Both are plotted against shoal area in Fig.~\ref{fig:branching}A. The mean latency to the first repeat wave, over the $n = 125$ attacks carrying a valid timing measurement, is $\bar{T}_{(1)} = 4.92$~s.

The pooling count $M$, the number of effectively independent responders whose earliest crossing produces the observed group response, is identified by the shape of the latency distribution rather than assumed. Both the inverse-Gaussian shape parameter $\lambda$ and the mean carry units of time, so neither alone constrains $M$, but the scale-free ratio $\lambda/\bar{T}_{(1)}$ rises steeply with the number of pooled accumulators. The threshold is not free either, since naming $M$ fixes it at $\theta_1(M) = -\log[1-(1-q)^{1/M}]$ through the measured false alarm rate, leaving a one-parameter family in $M$ alone. Against a Monte Carlo calibration of the minimum of $M$ first passages, the observed $\lambda/\bar{T}_{(1)} = 3.93$ gives
\begin{align}
  \hat{M} = 13.5 \quad (95\%\ \text{CI}\ [8.8,\,29.4]),
  \label{eq:Mfit}
\end{align}
plotted against the calibration curve in Fig.~\ref{fig:branching}B. Substituting the two response rates into the identification at this pooling count gives
\begin{align}
  \hat\alpha = 0.95 \quad (95\%\ \text{CI}\ [0.91,\,0.98]),
  \label{eq:fitresult}
\end{align}
where the interval is a joint bootstrap propagating both rate sampling error and the uncertainty in $\hat{M}$, resampled over recording-by-bout clusters and stratified by event type (Fig.~\ref{fig:branching}C). The implied per-unit false-alarm rate is $q_1 = 0.028$, so $\theta_1 = -\log q_1 = 3.57$ and the admissibility ceiling $K_{\max} = \log(1-q_1)/\log(1-q_1^2) = 36$ lies well above any plausible attended neighborhood.

Reading $\bar{T}_{(1)}$ as a group minimum rather than a single first passage makes $M$ identifiable from the data. A single-passage inverse Gaussian would require $\theta \approx 154$ to reproduce the observed shape, more than forty times the threshold implied by the rates. The latencies also show no censoring signature, with a smoothly decaying tail ($95$th percentile $9.1$~s, maximum $27.8$~s, a single event within $2$~s of the maximum) rather than the accumulation a fixed observation window would leave, so the $50$ unanswered attacks are misses rather than truncated late responses.

The estimate is well above the individually Bayesian benchmark. Since $L_\infty$ increases in $K$ and $K \le M$, the largest value the benchmark can take at a given pooling count is $L_\infty(M)$, and $\hat\alpha$ exceeds even that in every bootstrap replicate, with a minimum excess of $0.20$ and $P(\hat\alpha > L_\infty) = 1.000$ (Fig.~\ref{fig:branching}C). Because $\hat\alpha$ depends on $M$ alone and the benchmark on $K$ alone, the conclusion does not require $K$ to be identified. Heterogeneity in $K$ across individuals widens rather than closes the gap, since $L_\infty$ is concave over this range and $\E[L_\infty(K)] < L_\infty(\E[K])$ (\figS{4C}). These fish therefore discount social evidence more steeply than Bayesian updating over any admissible neighborhood would supply.

We address two caveats to estimating $\hat{M}$ that propagate to $\hat\alpha$. The calibration treats the $M$ accumulators as independent, but correlated or occluded visual fields would shrink the effective count, moving $\hat{M}$ down and $\hat\alpha$ with it (\figS{3C}). Also, the false-alarm rate rests on only $26$ independent recording-by-bout clusters, so the independent-events interval is not a reliable guide to the uncertainty in $q$, and the clustered intervals we report differ from it in both directions, up to $24\%$ wider for $\mathrm{TP}$ and $21\%$ narrower for $q$ (\figS{5B}). However, neither caveat threatens the sign of the result, since $\hat\alpha$ exceeds $L_\infty(M)$ at every $M$ in the bootstrap range (\figS{4B}).

\subsubsection*{A false alarm rate that does not grow with the shoal}
\label{sec:scaling}

The fit used two pooled rates and a latency distribution, all measured at a single shoal size. How those rates should \emph{scale} with shoal size is not something the fit constrains, and the model makes a sharp prediction about one of them.

The threshold scaling of Eq.~(\ref{eq:threshold_scaling}) exists precisely to cancel the $K$-fold amplification of individual false alarms, fixing the group false alarm rate at its solitary value independent of shoal size and of the discounting rate. The naive model has no such cancellation, since at fixed threshold $q_N \approx N q_\alpha(\theta)$ grows with the number of fish that could trigger a wave, so a naive shoal should produce false alarms more often as it grows. The two models therefore differ in sign on an observable neither was fit to.

The data are flat. Fitting a logistic in shoal area to the flyby responses gives a slope of $-0.006$ (SE $0.021$, $P = 0.78$) across areas from $31$ to $75$~m$^2$, consistent with a constant rate and inconsistent with the increase the naive model requires (Fig.~\ref{fig:branching}A). Over the same range the true-positive rate rises, $+0.033$ (SE $0.014$, $P = 0.019$), so detection improves with shoal size while the error rate does not. These standard errors treat events as independent; the clustered versions are reported in \figS{5B} and do not change either sign. A pooled model with an event-type interaction does not resolve the two slopes apart ($z = 1.54$, $P = 0.12$). Each slope is therefore supported on its own, one flat and one rising, but the contrast between them is not, and we do not claim the two rates diverge.

This is the asymmetry the three counts predict. False alarms are controlled locally, through a threshold set by the attended count $K$, which does not grow when the shoal does. Detections are pooled globally, over every fish that might be first to respond, and that pool does grow. A collective can therefore become more sensitive without becoming more error-prone, which is the behavior the field study reports~\cite{pacher2025better} and which no model that ignores the evidence in a neighbor's stillness produces.

\section*{Discussion}
\label{sec:discussion}
 
The central contribution of this work is to describe and validate a mechanism of social discounting in collective escape, in which a rate $\alpha$ sets how strongly an animal discounts the evidence implied by neighbors who have not yet fled. The naive limit $\alpha=0$ ignores that evidence and the large-$K$ Bayesian limit $\alpha=1$ saturates its effects, with intermediate values interpolating as a principled, analytically tractable family. We showed that $\alpha$ is identified in closed form from two group-level observables, the true-positive and false-alarm response rates, in a ratio from which the threshold and the noise scale cancel, and that the count those rates pool over can be read off the escape latency shape rather than assumed. Fitting wild sulphur molly shoals gives $\hat\alpha = 0.95$ at a pooling count of $13.5$, above what individually Bayesian updating supplies over every neighborhood these fish could attend. The finding is therefore not that they approximate Bayesian social inference but that they discount a still neighbor more steeply than it prescribes, and the same model at that value predicts a false alarm rate constant in shoal size, a scaling law that played no role in the fit. Better and faster collective decisions thus emerge from individually rational accumulation of private and social evidence rather than from group-level coordination.

The heuristic also yields exact cascade statistics connecting the model to empirical work on subcritical escape waves. At the scaled threshold $\theta_K = (\theta_1 + \log K)/(1+\alpha)$, one departure triggers $b_{H=1} = K$ others on average under a real threat, while the scaling cancels the amplification exactly for false alarms and leaves $b_{H=0} = K q_\alpha(\theta_K) = q_1$, the solitary false alarm probability, at every group size. The collective amplifies real signals and damps spurious ones, with discriminability $b_{H=1}/b_{H=0} = K/q_1$ growing with the attended count. Because the false-alarm rate is anchored at the $K=1$ value it carries no dependence on $\alpha$, so discounting buys not a larger discriminability at fixed threshold but a lower threshold at matched accuracy, and hence faster detection. Supercritical spreading under threat and strongly subcritical cascades in safety thus follow from individually rational threshold setting rather than group-level tuning.

This separation gives a normative account of findings that phenomenological cascade models describe but do not derive, and it orders otherwise disparate systems on a single subcritical-to-critical axis. Startle cascades in golden shiner schools are subcritical at baseline and approach criticality in the alarm state~\cite{poel2022subcritical}, a pattern read as managing a sensitivity-robustness tradeoff but left without mechanistic origin. In our framework $b_{H=0}$ is fixed at its single-agent value regardless of $K$, and alarm-state shifts correspond to a lowered threshold that raises $q_1$ and drives $b_{H=0}$ toward criticality with $b_{H=1} = K$ unchanged. A reef-fish system governed instead by response rescaling, a gain-control rule rather than evidence accumulation~\cite{fahimipour2023misinformation}, sits near $b \approx 1$, showing the same balance is reachable without discounting, while excitable dynamics in sulphur molly shoals place them at the strongly subcritical end~\cite{gomeznava2023excitable}, consistent with the large $\hat\alpha$ we recover. Social discounting thus supplies both a mechanism and a single ordering parameter for the alarm-state modulation that branching statistics describe but cannot explain.

The molly data are the source of our fit rather than an independent test of it~\cite{pacher2025better}, though how the fitted quantities scale with shoal size is not something the fit constrains. Across 177 kiskadee attacks and 81 harmless flybys the true-positive rate rose with shoal size while the false alarm rate stayed essentially constant, the pattern a threshold scaling with group size predicts and the decoupling of true and false positives anticipated by earlier collective-decision theory~\cite{wolf2013accurate}. In our framework the two come from different counts, false alarms controlled locally through a threshold set by the attended count and detections pooled over every fish that might be first to respond. The reported decrease of detection time with shoal size is consistent with this but is not a further test, since the latency shape is what identifies the pooling count. Nor is the one quantity the fit does not address, the dependence of latency on shoal area, since its log-log slope is negative as a first-passage minimum requires but the clustered interval crosses zero, and the companion structural dataset shows that area does not track headcount within a site, so it cannot stand in for pool size (\figsS{6A, S6B}). Similar joint gains in speed and accuracy appear in the laboratory, where solitary fish avoided a replica predator in roughly half of trials while groups of sixteen approached near-perfect accuracy~\cite{ward2011fast}, and the self-organized division of vigilance proposed for that result~\cite{ward2011fast, ward2008quorum, sumpter2009quorum} is a behavioral implementation of the private evidence accumulation in our model.

Group-size effects on escape thresholds have a long history in behavioral ecology. Animals in groups tolerate closer predator approach before fleeing than solitary individuals~\cite{lima1998stress}, an effect on flight initiation distance documented across taxa~\cite{ydenberg1986fleeing, cooper2015escaping}. Our threshold scaling supplies a normative derivation, since larger groups should set higher individual thresholds not because the predator is less dangerous but because the collective false alarm rate amplifies with $K$. The flight initiation framework, in which escape balances the cost of fleeing too early against the risk of fleeing too late~\cite{ydenberg1986fleeing, cooper2015escaping, cresswell2000evidence}, maps onto our speed-accuracy frontier with the threshold playing the role of the critical approach distance, and the dependence of its slope on $\alpha$, from $1$ in the naive limit to $\tfrac12$ in the Bayesian limit, makes flight initiation distance a potential behavioral readout of the discounting rate.

Escape waves are initiated by a few highly responsive individuals and spread by neighbor copying, with the two governed by distinct mechanisms~\cite{herbert2015initiation}, and our framework accounts for both. Initiation rates are observable through the statistics of the minimum first-passage time over the detecting pool, driven by the agent whose private evidence first crosses through an unlikely fluctuation rather than by a systematically lower threshold, giving a detection time falling logarithmically with pool size. Spreading is governed by $b_{H=1} = K \gg 1$, which guarantees a complete cascade, and $b_{H=0} = q_1 \ll 1$, which ensures it dies out, consistent with waves passing through entire groups during real threats~\cite{herbert2015initiation} and with subcritical cascade sizes at baseline~\cite{poel2022subcritical}. Rapid waves after robotic-falcon attack in starling and pigeon flocks, quickly dying in safe conditions~\cite{papadopoulou2022selforganization, storms2024robotic}, support the generality of this rate separation across species.

This normative derivation complements mechanistic accounts based on spatial self-organization~\cite{klamser2021collective}, and both can operate at once, with social discounting fixing the false alarm rate across group sizes while school geometry governs the spatial spread of the wave. That denser schools sit closer to criticality~\cite{poel2022subcritical, klamser2021collective} is then density modulating the effective neighbor count, since denser schools present more visible neighbors and a threshold not readjusted to the new count lets $b_{H=0}$ rise toward criticality. This predicts that manipulating density while holding $K$ fixed, achievable by controlling visual access, leaves $b_{H=0}$ unchanged, whereas raising $K$ at fixed density pushes it toward criticality. Sparse, speed-dependent interaction networks~\cite{rosenthal2015revealing, strandburg2013visual} fit naturally, since coupling with effective degree $K \ll N$ fixes the group false alarm rate and the logarithmic threshold scaling, and the private evidence $\xi_i(t)$ and survival log-likelihood ratio $\Lambda^{\rm surv}(t)$ map onto the individual and collective components of risk encoding identified in golden shiner schools~\cite{sosna2019individual}.

Whether a given collective is better described by the naive or the Bayesian model is a question about the computational sophistication of its members. The naive rule needs only a fixed belief increment when a neighbor departs, implementable by simple circuits with no representation of survival timing~\cite{yilmaz2013rapid, evans2019cognitive}, whereas the Bayesian rule requires a continuous computation of the survival log-likelihood ratio, which the heuristic approximates with a single constant anti-drift. The coarse behavioral signature, a false alarm rate that does not grow with the shoal, holds for any group that scales its threshold with the attended count, so it cannot by itself separate a strong discounter from a partial one or from a threshold-scaling naive group. What pins $\alpha$ down is the identification from the two response rates together with the pooling count, and at finer resolution the departure-jump distribution, a fixed increment in the naive model but growing as $\theta - \Lambda^{\rm surv}(T_j)$ in the Bayesian one, resolvable by tracking individual belief trajectories around a neighbor's departure as high-resolution behavioral tracking improves~\cite{rosenthal2015revealing, sosna2019individual}.

Three predictions distinguish this framework from phenomenological branching models and are testable with existing methods. The false alarm branching ratio should satisfy $b_{H=0} \approx q_1$ independent of total group size at the scaled threshold; the true-threat ratio should satisfy $b_{H=1} \approx K$, so their ratio measures collective discriminability and through it the discounting rate; and the threshold should grow as $\log K/(1+\alpha)$, testable by manipulating group size and visual access independently. None has been measured directly, since the molly response rates are consistent with a fixed group false alarm rate but are not measurements of either branching ratio, and published branching ratios for schooling fish~\cite{poel2022subcritical} have not been analyzed against the scaled-threshold prediction. Each would supply an independent estimate of $\alpha$ to set against the $\hat\alpha = 0.95$ obtained here.

We envision several interesting extensions to this work. The locally coupled mean-field treatment is tractable but abstracts away the spatial structure through which escape waves actually propagate~\cite{herbert2015initiation, papadopoulou2022selforganization}. Replacing the fixed neighborhood with a spatial interaction kernel on a ring or plane would let us predict the speed and spatial extent of a wave from the range of that kernel. Heterogeneous thresholds, arising from variation in boldness or responsiveness~\cite{sosna2019individual}, would change how false alarms scale with group size once the threshold distribution has nonzero variance. What sets the pooling count is itself open, since we recover $\hat M = 13.5$ from a shoal holding on the order of $10^5$ fish, and whether that reflects visual occlusion, a limited zone of influence, or the geometry of wave initiation is not something the response rates can settle. Finally, a non-stationary environment in which the threat state changes over time would call for adaptive discounting that leaks evidence at a rate set by volatility, connecting social discounting to evidence accumulation in switching environments~\cite{veliz2016stochastic, glaze2015normative}, a computation animals perform near-optimally even when evidence arrives in discrete pulses~\cite{piet2018rats}, as neighbor departures do here.

\section*{Materials and Methods}
\label{sec:methods}

Each agent maintains an internal log-likelihood ratio $\xi_i(t)$ representing the private-evidence component of its belief about the threat state.
In the absence of social signals, $\xi(t)$ evolves as a drift-diffusion process
\begin{align}
  \dd\xi = \mu_H\,\dd t + \sigma\,\dd W_t, \qquad \xi(0) = 0,
  \label{eq:ddm_model}
\end{align}
where $W_t$ is a standard Wiener process, $\mu_H$ is the drift under hypothesis $H$, and $\sigma > 0$ is the noise amplitude.
Because $\xi$ is a log-likelihood ratio, its drift magnitude is half its variance rate~\cite{bogacz2006physics, gold2002banburismus}, leaving a single free scale that we absorb into the unit of time,
\begin{align}
  \mu_1 = 1, \qquad \mu_0 = -1, \qquad \sigma = \sqrt{2}.
  \label{eq:symmetric}
\end{align}
Thresholds are then measured in nats of log-odds. An agent departs at the first passage time
\begin{align}
  T = \inf\{t \geq 0 : \xi(t) = \theta\},
  \label{eq:fpt_model}
\end{align}
where $\theta > 0$ is its departure threshold.

\subsection*{Single-agent first passage}
\label{sec:methods_fpt}

Under the symmetric parametrization~\eqref{eq:symmetric}, the first-passage density is the inverse Gaussian~\cite{siegert1951first, redner2001guide}
\begin{align}
  f(t\mid H) = \frac{\theta}{\sqrt{4\pi t^3}} \exp\!\left[-\frac{(\theta - \mu_H t)^2}{4t}\right],
  \label{eq:igdensity}
\end{align}
and the survival function is
\begin{align}
  S(t\mid H) = \Phi\!\left(\frac{\theta - \mu_H t}{\sqrt{2t}}\right) - e^{\mu_H\theta}\,\Phi\!\left(\frac{-\theta - \mu_H t}{\sqrt{2t}}\right),
  \label{eq:survival}
\end{align}
where $\Phi$ is the standard normal CDF.
When $H=1$, the drift carries the particle to threshold with probability one, and the first moment of~\eqref{eq:igdensity} is the mean detection time $\bar{T} \equiv \E[T\mid H=1] = \theta/\mu_1 = \theta$.
When $H=0$, the particle drifts away from threshold at unit rate with diffusivity $D = \sigma^2/2 = 1$, so crossings are fluctuation-driven, and taking $t \to \infty$ in~\eqref{eq:survival} leaves the false alarm probability
\begin{align}
  q(\theta) := \PP(T < \infty \mid H=0) = e^{-\theta},
  \label{eq:fap}
\end{align}
the escape probability $e^{-\mu_1 \theta / D}$ against a drift.

The hazard rate $h_H(t) = f(t\mid H)/S(t\mid H)$ is the instantaneous departure rate given survival to $t$. The log-likelihood ratio of survival,
\begin{align}
  \Lambda^{\mathrm{surv}}(t) = \log\frac{S(t\mid H=1)}{S(t\mid H=0)} \le 0,
  \label{eq:survllr}
\end{align}
quantifies the negative evidence in continued non-departure; it is zero at $t=0$ and grows in magnitude through the period when departure would be expected for $H=1$ (Fig.~\ref{fig:singleagent}B). Its rate of accumulation is $h_0(t)-h_1(t)$, the negative of the single-agent hazard gap $\lambda_{\rm single}(t)=h_1(t)-h_0(t)$.

\subsection*{Social coupling: naive and Bayesian models}
\label{sec:methods_coupling}

When agent $j$ departs at $T_j$, its neighbors update their beliefs, and the two models differ in whether the silence preceding that departure is treated as informative.
The \emph{naive pulsatile model} discards the timing and credits each departure with a fixed positive increment $\kappa > 0$,
\begin{align}
  \dd\xi_i = \mu_H\,\dd t + \sigma\,\dd W_i(t) + \kappa \sum_{j \in \mathcal{N}(i)} \delta(t - T_j)\,\dd t,
  \label{eq:naive_model}
\end{align}
where $\mathcal{N}(i)$ is the set of agents observed by $i$.
The \emph{Bayesian model} uses the whole of $j$'s trajectory, so $j$ contributes the survival evidence $\Lambda^{\mathrm{surv}}$ of Eq.~\eqref{eq:survllr} while inactive and the log density ratio once it goes,
\begin{align}
  \xi_i^{\mathrm{full}}(t) = \xi_i(t) + \sum_{j \in \mathcal{N}(i)}
  \begin{cases}
    \Lambda^{\mathrm{surv}}(t), & t < T_j, \\[4pt]
    \Lambda^{\mathrm{surv}}(T_j) + \log\dfrac{h_1(T_j)}{h_0(T_j)}, & t \geq T_j.
  \end{cases}
  \label{eq:bayesian_model}
\end{align}
A neighbor's silence therefore drives a continuous downward drift, and its departure produces an upward jump equal to the log hazard ratio at $T_j$, which grows with $T_j$ and so offsets the negative evidence accumulated while $j$ was still. For the dyad, the coupled belief processes evolve as
\begin{align}
  \dd\xi_i = \mu_H\,\dd t + \sigma\,\dd W_i(t) - \lambda(t)\,\dd t + \left[\theta - \Lambda^{\rm surv}(T_j)\right]\delta(t - T_j),
  \label{eq:bayesian_sde}
\end{align}
where the survival correction $\lambda(t) = -\tfrac{\dd}{\dd t}\Lambda^{\mathrm{surv}}(t) = h_1(t) - h_0(t)$ is determined self-consistently. Each agent drifts at $\mu_H - \lambda(t)$ while its neighbor is silent, the hazards $h_H$ are the first-passage hazards at that drift, and the loop closes through $\lambda(t) = h_1[\lambda](t) - h_0[\lambda](t)$. The departure jump $\theta - \Lambda^{\rm surv}(T_j)$ is derived in Supplementary Materials; since $\Lambda^{\rm surv}\le0$ it always exceeds $\theta$, encoding that a neighbor surviving long before departing provides stronger evidence for a real threat.

The neighbor sets $\mathcal{N}(i)$ define the topology. We analyze an \emph{all-to-all} network, $\mathcal{N}(i) = \{1,\ldots,N\}\setminus\{i\}$, yielding exact false-alarm and threshold-scaling results, and a \emph{random neighbor} model in which each departing agent is observed by a randomly selected subset of $K$ others, capturing the sparse, locally concentrated interaction structure observed empirically~\cite{rosenthal2015revealing} and yielding the cascade branching ratios. Because every all-to-all result depends on group size only through the number of agents an individual attends, each carries over to the sparse case with $N$ replaced by $K$; the empirical identification uses that form throughout and at no point assumes all-to-all connectivity in the shoal.

\subsection*{The heuristic linear anti-drift model}
\label{sec:methods_heuristic}

The Bayesian survival correction $\lambda^{(N)}(t)$ has no closed form at any $N$, including the dyad, since each agent's discount raises the effective boundary that determines the very hazards the discount is built from.
It is fixed implicitly by that self-consistent first-passage problem, one solve per group size, which we carry out by damped fixed-point iteration for $N$ up to $100$ (Sec.~\ref{sec:fixedpoint} and Fig.~\ref{fig:largeN}A, with the boundary, kernel, and closure given in Sec.~\ref{sec:durbin}).
The Bayesian correction rises to a bounded plateau, providing an asymptotically constant anti-drift. A neighbor's silence is negative evidence accumulating at a saturating rate. A constant discounting rate is therefore well-motivated and tractable, providing appropriate threshold scalings and the frontier in closed form.
We therefore also consider a model that replaces $\lambda^{(N)}(t)$ with a constant that can be tuned to the Bayesian model's long-time limit or other values $\alpha \in [0,1]$.
During a neighbor's silence each agent carries a constant downward drift $-\alpha$, with a time-dependent kick on departure,
\begin{align}
  \dd\xi_i = (\mu_H - \alpha)\,\dd t + \sigma\,\dd W_i(t) + \sum_j \left[\theta + \alpha T_j\right]\dd N_j(t),
  \label{eq:heuristic_sde}
\end{align}
where $N_j(t)$ counts neighbor $j$'s departures.

The silence-phase drifts are $\tilde\mu_1 = 1-\alpha$ and $\tilde\mu_0 = -(1+\alpha)$, and the kick is the constant-rate form of the exact Bayesian update $\theta-\Lambda^{\rm surv}(T_j)$ for $\Lambda^{\rm surv}(t)\approx-\alpha t$. The model is fixed by $(\alpha,\theta)$, and its single-agent statistics are the standard first-passage results at the shifted drifts,
\begin{align}
  q_\alpha(\theta) = e^{-(1+\alpha)\theta}, \qquad \bar{T}_\alpha(\theta) = \frac{\theta}{1-\alpha}.
  \label{eq:heuristic_observables}
\end{align}

\paragraph{Interpretation of $\alpha$.} The rate $\alpha$ sets how strongly an agent weights the negative evidence in a neighbor's continued silence. At $\alpha=0$ that evidence is discarded and the rule reduces exactly to the naive pulsatile model, with kick $\kappa=\theta$ and no anti-drift, while larger $\alpha$ uses more of the evidence a Bayesian agent would use, which is not the same as a closer pointwise match to the Bayesian belief trajectory. We restrict to $\alpha\in[0,1]$ on behavioral grounds, since $\tilde\mu_1$ vanishes at $\alpha=1$ and reverses beyond it, letting a neighbor's silence drive a genuine detector away from threshold.

\paragraph{Required discounting rate.} A group of $K$ attending agents holds its false alarm rate at the solitary value $q_1$ when $1-(1-q_\alpha)^K = q_1$, so at the threshold $\theta_1 = -\log q_1$ the required rate is
\begin{align}
  \alpha_{\rm req}(K, q_1) = \frac{\log\!\left[1/\left(1-(1-q_1)^{1/K}\right)\right]}{\log(1/q_1)} - 1,
  \label{eq:alpha_req}
\end{align}
zero at $K=1$ and rising as $\log K/\log(1/q_1)$ thereafter (Fig.~\ref{fig:largeN}D), fixing $\theta_1$ by observable false alarm rates from data. Admissibility caps the attended count at $K_{\max}(q_1) = \log(1-q_1)/\log(1-q_1^2) \simeq 1/q_1 \approx 36$ at the fitted $q_1 = 0.028$, so the molly operating point at $K=7$ is bound by the comparison with $L_\infty(K)$ rather than by admissibility.

\paragraph{Group first-departure time.} The group first-departure time $T_{(1)} = \min_i T_i$ has density $f_{(1)}(t \mid H) = N f(t \mid H)\,S(t \mid H)^{N-1}$, and for large $N$ its conditional mean obeys the saddle-point scaling
\begin{align}
  \bar{T}_{(1)} \approx \frac{\theta^2}{2 \sigma^2 \log N},
  \label{eq:group_T1}
\end{align}
accurate to about ten percent at $N=100$. The discounting rate does not appear at leading order, entering only through $\theta(\alpha)$ and as an additive shift inside the logarithm. All reported detection times are computed exactly from $\bar T_{(1)} = \int S(t \mid H{=}1)^N\,\dd t$.

\paragraph{Derivation of the saturated drift.} Write $L^{(N)} \equiv (N-1)\lambda^{(N)}$ for the total Bayesian survival drift, so that each agent's effective drift in response to silent neighbors approaches $1 - L_\infty$ for $H=1$ and $-1 - L_\infty$ for $H=0$. For constant drift $\nu$ and $\sigma^2 = 2$ against a fixed barrier, the first-passage hazard tends to $\nu^2/4$ for $\nu > 0$ and to zero otherwise, since a survivor pushed away from threshold crosses at a vanishing rate. The $H=0$ drift is negative for any $L_\infty \ge 0$, so the false-alarm hazard vanishes asymptotically and the closure $\lambda^{(N)} = h_1 - h_0$ reduces in the tail to $L_\infty = (N-1)(1-L_\infty)^2/4$. We take the root in $(0,1)$, Eq.~(\ref{eq:Linf}); the other exceeds unity and would reverse the sign of the $H=1$ drift. At fixed $N$, $L^{(N)}(t)$ decays toward its late-time value from above, so that value is obtained by tail extrapolation and a mean over the decision window overstates it.

\paragraph{Relation to the Bayesian model.} Three statements connect the heuristic to the full Bayesian analysis, at different levels of rigor, and we keep them distinct. (i)~\emph{Exact.} The total correction has the bounded closed-form limit $L_\infty(N) = (\sqrt N - 1)/(\sqrt N + 1)$ of Eq.~\eqref{eq:Linf}, rising monotonically toward unity as $1 - 2/(\sqrt N + 1)$, which is what makes a constant discount in $[0,1]$ a reasonable summary and fixes the scale of the upper endpoint. (ii)~\emph{Verified numerically at the dyad.} Setting $\alpha=L_\infty(2)$, the heuristic reproduces the $N=2$ first-departure time to $1.5\%$ relative error, against $2.6\%$ for a decision-window plateau mean, the two bracketing the exact value from below and above, and its survival functions match the full Bayesian dyad to $2.6\times10^{-2}$ for $H=1$ and $1.1\times10^{-3}$ for $H=0$ (Fig.~\ref{fig:dyad}E). These are $N=2$ figures and do not transfer, since the $H=1$ discrepancy grows with group size for the reason given in Sec.~\ref{sec:fixedpoint} while the $H=0$ discrepancy stays of order $10^{-3}$ at every $N$. (iii)~\emph{Approximate, increasingly so with $N$.} The constant-$\alpha$ rule does not track the time-varying correction pointwise, which rises through an early transient exceeding unity at large $N$ before relaxing (Supplementary Materials), and the constant-drift idealization degrades as the residual slope of $L^{(N)}$ over the decision window grows (\figS{2}). Nothing below depends on that pointwise match. The empirical results need only that $\alpha$ is identified from data and that a group scaling its threshold with the attended count holds its false alarm rate flat as it grows, while one holding its threshold fixed amplifies.

\subsection*{Numerical solution of the coupled first-passage problem}
\label{sec:methods_numerics}

The coupled survival functions require first-passage statistics of a drift-diffusion with time-varying drift $\mu_H - (N-1)\lambda(t)$ against the fixed threshold $\theta$. A deterministic shift $y(t)=\xi(t)-\int_0^t(\mu_H-(N-1)\lambda)\,\dd s$ converts this to a driftless process against the moving boundary $b(t) = \theta - \int_0^t(\mu_H-(N-1)\lambda)\,\dd s$, whose first-passage density solves the second-kind Volterra integral equation of Durbin~\cite{durbin1971boundary, durbin1985first} in the master-equation form of Buonocore, Nobile, and Ricciardi~\cite{buonocore1987new}. This is solved forward in physical time by a single lower-triangular quadrature sweep, with no spatial grid and no linear solve, identically for both hypotheses; the self-consistent $\lambda(t)$ is obtained by damped fixed-point iteration. The full method, and its advantages over a backward-Kolmogorov solver, are given in Section~\ref{sec:durbin}, where the solver is checked against the exact inverse Gaussian on both a constant and a linear moving boundary. The dyad cascade probabilities are checked separately against the closed forms of Eqs.~(\ref{eq:Gexact})--(\ref{eq:Cexact}) in Section~\ref{sec:orderstat}. Where trajectories are simulated rather than solved, we integrate by Euler--Maruyama and shift the barrier inward by $0.5826\,\sigma\sqrt{\Delta t}$ to correct discrete-barrier bias~\cite{broadie1997continuity}; uncorrected, the bias inflates the simulated dyad cascade probability by about ten percent, comparable in size and sign to the Bayesian--naive offset in Fig.~\ref{fig:dyad}F.

\subsection*{Cascade size and discriminability}
\label{sec:methods_cascade}

The full expected cascade size, including the initiating agent and weighted by the probability of initiation for $H=0$, is
\begin{align}
  \E[\text{cascade}\mid H=1] &= 1 + K\,P\!\left(\xi_i(T_{(1)}^-) \geq 0 \mid H=1\right), \nonumber \\
  \E[\text{cascade}\mid H=0] &= \Bigl[1 + K\,P\!\left(\xi_i(T_{(1)}^-) \geq 0 \mid H=0\right)\Bigr]\,P\!\left(T_{(1)} < \infty \mid H=0\right),
  \label{eq:cascade_sizes}
\end{align}
where $K$ is the number of agents attending a given individual and $P(\xi_i(T_{(1)}^-)\geq0)$ is the fraction of them whose pre-kick belief lies above zero. We evaluate this by integrating each agent's killed belief distribution at $T_{(1)}$ against the distribution of $T_{(1)}$, giving the finite-$K$ discriminability $d = \E[\text{cascade}\mid H=1]/\E[\text{cascade}\mid H=0]$. Complete clearing for $H=1$ gives cascade size $K+1$ and branching ratio $b_{H=1} = K$, the convention used throughout.

\paragraph{The dyad frontier.} The frontier of Fig.~\ref{fig:dyad}F is swept over $\theta$ at each of $22$ values of $\alpha$, with the naive rule recovered at $\alpha=0$, and the tuned envelope taken as the lower hull over $\alpha$ at matched $\bar T_{(2)}$. That envelope improves on the naive rule by at most $0.6\%$ in cascade rate, and $\alpha=0$ is itself the minimizing member over much of the range. The exact Bayesian dyad is evaluated by Monte Carlo at seven thresholds, $4\times10^4$ trials per hypothesis, with a Broadie--Glasserman--Kou continuity correction shifting the simulated barrier inward by $0.5826\,\sigma\sqrt{\Delta t}$ to remove the discrete-monitoring bias~\cite{broadie1997continuity}. Its cascade rate sits $4$ to $17\%$ above the naive curve at matched response time, resolved at $3$ to $5$ Monte Carlo standard errors at the lower thresholds.

\subsection*{Empirical identification}
\label{sec:methods_identification}

Pooling over the $M$ responders whose earliest crossing registers as a group response, the per-unit response probabilities are recovered from the group rates by $q_{\rm ind} = 1-(1-q)^{1/M}$ and $p_{\rm ind} = 1-(1-\mathrm{TP})^{1/M}$, with $\mathrm{miss}_{\rm ind} = 1-p_{\rm ind}$, where $q$ is the group false-alarm rate and $\mathrm{TP}$ the group true-positive rate. In the heuristic model these are the two threshold-crossing probabilities under the shifted drifts,
\begin{align}
  q_{\rm ind} = e^{-2(1+\alpha)\theta/\sigma^2}, \qquad
  \mathrm{miss}_{\rm ind} = e^{-2(1-\alpha)\theta/\sigma^2},
  \label{eq:cross}
\end{align}
the nonzero miss arising because the collective response is registered only within a finite decision window. The ratio of their logarithms cancels the common factor $2\theta/\sigma^2$, giving the SNR-free estimator $\hat\alpha = (r-1)/(r+1)$ with $r = \log q_{\rm ind}/\log\mathrm{miss}_{\rm ind} = (1+\alpha)/(1-\alpha)$. The full estimators, the data source (Pacher et al.\ Data~S1~\cite{pacher2025better}), the bootstrap confidence intervals, the area-scaling model fits, and the identification of the pooling count from the latency shape are detailed in Supplementary Materials. The pooling count $M$ is identified from the latency shape and is not fixed by hand; the attended count $K$ is not identified, and the over-discounting comparison is constructed so that it need not be, since $L_\infty$ is increasing and $K \le M$ bounds the benchmark by $L_\infty(M)$.


\paragraph{Funding.}
This research was supported in part by grants from the NSF (DMS-2235451) and the Simons Foundation (MPS-NITMB-00005320) to the NSF-Simons National Institute for Theory and Mathematics in Biology (NITMB), and by the NSF (DMS-2527338 \& IIS-2616531).

\paragraph{Author contributions.}
Z.P.K. conceived the study, developed the theory and numerical methods, performed the analysis, and wrote the manuscript.

\paragraph{Competing interests.}
The author declares no competing interests.

\paragraph{Data and materials availability.}
All data needed to evaluate the conclusions are present in the paper, the Supplementary Materials, and the cited source dataset. This study reanalyzes the publicly archived field dataset of Pacher et al.~\cite{pacher2025better}. All code implementing the drift-diffusion model, the moving-boundary first-passage solver, the fixed-point iteration, the Monte Carlo cross-checks, and the empirical identification is archived at Zenodo~\cite{kilpatrick2026code} and developed at \url{https://github.com/zpkilpat/collective-escape}.

\clearpage
\appendix
\renewcommand{\thesection}{S\arabic{section}}
\renewcommand{\thefigure}{S\arabic{figure}}
\renewcommand{\theequation}{S\arabic{equation}}
\renewcommand{\thesubsection}{S\arabic{section}.\arabic{subsection}}
\setcounter{section}{0}
\setcounter{figure}{0}
\setcounter{equation}{0}


\begin{center}
{\large\bfseries\sffamily Supplementary Materials for}\\[0.6em]
{\large\bfseries\sffamily Social Discounting Enables Fast and Reliable Collective Escape}\\[0.8em]
{Zachary P.~Kilpatrick}\\[0.4em]
{\small Corresponding author: \texttt{zpkilpat@colorado.edu}}
\end{center}

\vspace{1em}
\noindent
{\bfseries\sffamily This PDF file includes:}
\begin{itemize}\itemsep0pt
  \item Supplementary Text, Sections S1 to S9
  \item Figs.~S1 to S6
\end{itemize}

\vspace{1em}
\noindent
{\bfseries\sffamily Supplementary Text}

\vspace{0.5em}
\noindent
Sections~\ref{sec:prelim} to~\ref{sec:jump_supp} collect the killed densities, the dyad cascade integrals, and the Bayesian departure-jump identity. Sections~\ref{sec:durbin} and~\ref{sec:fixedpoint} give the moving-boundary first-passage method used for every coupled survival function and the iteration that closes the Bayesian fixed point. Sections~\ref{sec:Linf_supp} and~\ref{sec:hump} derive the saturation ceiling $L_\infty(N)$ in closed form and resolve the early-time hump by matched asymptotics, establishing that the hump diverges while only the tail saturates. Section~\ref{sec:largeN_supp} derives the extreme-value detection speedup, the threshold scaling, and the branching-ratio pinning. Section~\ref{sec:identification_supp} gives the empirical identification in full, covering the data, the response rates, the estimator of $\alpha$, the identification of the pooling count from the latency shape, and the area scaling.

Throughout we use the symmetric parametrization $\mu_1 = 1$, $\mu_0 = -1$, $\sigma = \sqrt2$ (variance rate $\sigma^2 = 2$, diffusion coefficient $1$), so the private accumulator $\xi$ is itself a log-likelihood ratio.

\section{Killed densities by the method of images}
\label{sec:prelim}

The single-agent accumulator obeys $\dd\xi = \mu_H\,\dd t + \sigma\,\dd W_t$ with $\xi(0)=0$ and absorption at $\theta>0$, with mean detection time $\bar T = \theta$, false alarm probability $q(\theta) = e^{-\theta}$, and inverse-Gaussian density and survival function as given in Materials and Methods.

The cascade integrals require the sub-probability density of an agent's position conditioned on \emph{not} having been absorbed, the killed density $p_H(x,t)$ on $x<\theta$ with $p_H(\theta,t)=0$. The method of images places a negative source at the reflection $2\theta$ of the origin across the barrier, weighted by $e^{\mu_H\theta}$ so the combination vanishes on $x=\theta$,
\begin{align}
  p_0(x,t) &= \frac{1}{\sqrt{4\pi t}}\left[e^{-(x+t)^2/4t} - e^{-\theta}\,e^{-(x-2\theta+t)^2/4t}\right],
  \label{eq:images0_supp}\\
  p_1(x,t) &= \frac{1}{\sqrt{4\pi t}}\left[e^{-(x-t)^2/4t} - e^{\theta}\,e^{-(x-2\theta-t)^2/4t}\right],
  \label{eq:images1_supp}
\end{align}
for $x<\theta$, given $H=0$ and $H=1$ respectively. These are un-normalized, since integrating over $x<\theta$ returns $S(t\mid H)$ rather than unity, so the survival weight is carried inside the density and does not appear separately in the integrals below.

\section{Dyad order-statistic cascade integrals}
\label{sec:orderstat}

Let $T_{(1)}=\min(T_1,T_2)$. Because $T_1,T_2$ are i.i.d.\ before the first departure, $T_{(1)}$ has density $2 f(t\mid H) S(t\mid H)$ and each agent is equally likely to depart first.

\subsection{False alarm cascade probability}

Condition on the first departure at $t$ for $H=0$. The survivor sits at $\xi(t^-)=x$ distributed as $p_0(x,t)$ and receives the naive kick $\kappa$, then crosses $\theta$ with probability $e^{-\max(\theta-x-\kappa,0)}$. Integrating over $x<\theta$,
\begin{align}
  G(t;\kappa,\theta) = \int_{-\infty}^{\theta} p_0(x,t)\,e^{-\max(\theta-x-\kappa,\,0)}\,\dd x,
  \label{eq:Gdef_supp}
\end{align}
which evaluates to
\begin{align}
  G(t;\kappa,\theta) &= \Phi\!\left(\frac{\theta+t}{\sqrt{2t}}\right) - \Phi\!\left(\frac{\theta-\kappa+t}{\sqrt{2t}}\right) - e^{-\theta}\left[\Phi\!\left(\frac{-\theta+t}{\sqrt{2t}}\right) - \Phi\!\left(\frac{-\theta-\kappa+t}{\sqrt{2t}}\right)\right] \nonumber\\
  &\quad + e^{-(\theta-\kappa)}\Phi\!\left(\frac{\theta-\kappa-t}{\sqrt{2t}}\right) - e^{\kappa}\Phi\!\left(\frac{-\theta-\kappa-t}{\sqrt{2t}}\right).
  \label{eq:Gexact}
\end{align}
Summing the two orderings,
\begin{align}
  \PP(\text{cascade}\mid H=0) = 2\int_0^\infty f(t\mid H=0)\,G(t;\kappa,\theta)\,\dd t.
  \label{eq:both_supp}
\end{align}
At $\kappa=0$ this collapses to the independent baseline $q^2$; as $\kappa\to\infty$ it approaches $2q-q^2$.

\subsection{Cascade time}

When $H=1$, detection is certain, so the relevant quantity is $\E[T_{(2)}-T_{(1)}\mid H=1]$. Conditioned on the first departure at $t$ with the survivor at $x$, the residual crossing time is a point mass at zero with weight $\PP(x+\kappa\ge\theta)$ and otherwise the gap $\theta-(x+\kappa)$. Averaging against $p_1(x,t)$,
\begin{align}
  C(t;\kappa,\theta) &= (\theta-\kappa-t)\Phi\!\left(\frac{\theta-\kappa-t}{\sqrt{2t}}\right) + \sqrt{2t}\,\phi\!\left(\frac{\theta-\kappa-t}{\sqrt{2t}}\right) \nonumber\\
  &\quad - e^{\theta}\!\left[(-\theta-\kappa-t)\Phi\!\left(\frac{-\theta-\kappa-t}{\sqrt{2t}}\right) + \sqrt{2t}\,\phi\!\left(\frac{-\theta-\kappa-t}{\sqrt{2t}}\right)\right],
  \label{eq:Cexact}
\end{align}
and $\E[\text{cascade time}\mid H=1] = 2\int_0^\infty f(t\mid H=1)\,C(t;\kappa,\theta)\,\dd t$. As $\kappa\to\infty$ the atom dominates and the cascade time vanishes; at $\kappa=0$ it is the mean gap between the order statistics of two independent inverse Gaussians of mean $\theta$.

Both integrals are independent of the Volterra solver, so they check it: evaluating the dyad cascade probability from the converged survival functions and from Eqs.~(\ref{eq:Gexact}) and~(\ref{eq:Cexact}) agrees to four significant figures, giving $0.4008$, $0.1732$, $0.0647$, $0.0132$ at $\theta = 0.8$, $1.5$, $2.3$, $3.568$, each within its analytic bounds $[q^2,\,2q-q^2]$.

\section{The Bayesian departure-jump identity}
\label{sec:jump_supp}

When neighbor $j$ departs at $T_j$, two updates occur at once. Up to $T_j$ the observer has accumulated the survival evidence of $j$'s silence, a downward shift by $\Lambda^{\rm surv}(T_j) = \log[S(T_j\mid H{=}1)/S(T_j\mid H{=}0)] \le 0$. The departure itself carries the log density ratio, which for the inverse Gaussian is exactly $\theta$ whenever it occurs, since the prefactors cancel and the exponents differ by $[(\theta+t)^2-(\theta-t)^2]/4t$,
\begin{align}
  \log\frac{f(T_j\mid H=1)}{f(T_j\mid H=0)} = \theta.
  \label{eq:departure_llr}
\end{align}
Writing $f = hS$ splits that into the part already counted and the surprise of the crossing itself,
\begin{align}
  \log\frac{h_1(T_j)}{h_0(T_j)} = \theta - \Lambda^{\rm surv}(T_j) \ge \theta,
  \label{eq:jump_supp}
\end{align}
so the net update relative to the survival-shifted belief at $T_j^-$ is that hazard log-ratio, exceeding $\theta$ because $\Lambda^{\rm surv}\le0$. The constant-rate approximation $\Lambda^{\rm surv}(t)\approx-\alpha t$ turns it into the heuristic kick $\theta+\alpha T_j$ of Eq.~(\ref{eq:heuristic_sde}).

\section{Moving-boundary first passage by the Durbin--Buonocore method}
\label{sec:durbin}

The coupled survival functions solve the first-passage problem of a drift-diffusion with time-varying drift, $\dd\xi = (\mu_H - \Lambda(t))\,\dd t + \sigma\,\dd W_t$ with $\sigma^2=2$, against a fixed threshold, where $\Lambda(t) = (N-1)\lambda(t)$. After a deterministic shift this is a driftless process against a moving barrier: writing $A(t) = \int_0^t(\mu_H - \Lambda(s))\,\dd s$ and $y(t) = \xi(t) - A(t)$, absorption becomes first passage of $y$ to
\begin{align}
  b(t) = \theta - A(t).
  \label{eq:moving_boundary}
\end{align}
The first-passage density satisfies the Volterra equation of the second kind of Durbin~\cite{durbin1971boundary, durbin1985first} in the master form of Buonocore, Nobile, and Ricciardi~\cite{buonocore1987new},
\begin{align}
  g(t) = \Psi(t,0\mid b) - \int_0^t g(s)\,\Psi(t,s\mid b)\,\dd s,
  \label{eq:volterra}
\end{align}
where, with $p(y,t\mid y',s) = [4\pi(t-s)]^{-1/2}\exp[-(y-y')^2/4(t-s)]$,
\begin{align}
  \Psi(t,s\mid b) = \left[\frac{b(t)-b(s)}{t-s} - \dot b(t)\right] p\bigl(b(t),t\mid b(s),s\bigr).
  \label{eq:kernel}
\end{align}
Equation~(\ref{eq:volterra}) is solved by forward substitution on a uniform grid: $\Psi(t,s)$ is bounded as $s\to t$ so the diagonal is removable, and each $g(t_k)$ follows from $\{g(t_j)\}_{j<k}$ by one quadrature, giving the whole density in a single $O(n_t^2)$ pass with no linear solve. Then $S_H(t_k) = 1 - \Delta t\sum_{j\le k} g(t_j)$ and $h_H = g/S_H$.

Three properties of this method matter, the first for economy and the other two for correctness. The method is PDE-free, so there is no spatial grid and no spatial-discretization error, and accuracy is governed by the single step $\Delta t$; halving it moves the $N=2$ saturation tail from $0.1636$ to $0.1638$ and leaves $N=100$ unchanged at $0.8077$, which is why those recoveries are quoted to one significant figure. It integrates forward in physical time for both hypotheses through one code path, which removes a time-direction ambiguity where it would do most damage, since the late-time $H=0$ survival sets the hazard that fixes the saturation of Section~\ref{sec:Linf_supp}. And the sign and factor convention of Eq.~(\ref{eq:kernel}) reproduces the exact inverse Gaussian to machine precision on a constant boundary and on a linear moving boundary; the constant case alone cannot validate the kernel, since $\Psi(t,s)$ vanishes identically there for $s>0$ and only the $s=0$ term survives, so the linear case fixes the sign.

The dyad cascade integrals of Section~\ref{sec:orderstat} supply an independent check on the converged survival functions, and the closed form of Section~\ref{sec:Linf_supp} is recovered across the full range of $N$, so the saturation is a self-consistent hazard balance rather than a discretization artifact.

\section{Fixed-point iteration for the survival correction}
\label{sec:fixedpoint}

The per-neighbor correction is self-consistent. Each agent's effective drift while its $N-1$ neighbors are silent is $\mu_H - L(t)$ with $L = (N-1)\lambda$, the hazards are the first-passage hazards at that drift, and the correction must close $\lambda = h_1 - h_0$.

Two equivalent representations of that first-passage problem are useful. In forward form the killed density $p_H(x,t)$ on $x<\theta$ obeys the Fokker--Planck equation
\begin{align}
  \partial_t p_H = -\partial_x\bigl[(\mu_H - L(t))\,p_H\bigr] + \partial_x^2 p_H, \qquad p_H(\theta,t) = 0, \quad p_H(x,0) = \delta(x),
  \label{eq:fokkerplanck_supp}
\end{align}
with survival $S_H(t) = \int_{-\infty}^{\theta} p_H(x,t)\,\dd x$ and first-passage density the absorbing flux $f_H(t) = -\partial_x p_H|_{x=\theta}$, the diffusion coefficient being $\sigma^2/2 = 1$. In moving-boundary form the deterministic shift of Eq.~(\ref{eq:moving_boundary}) removes the drift and leaves a driftless process against $b(t) = \theta - A(t)$, whose first-passage density solves the Volterra equation~(\ref{eq:volterra}). The two describe the same object. We integrate the second, since the first requires a spatial grid, a truncated left boundary, and a derivative evaluated at the absorbing point, none of which the Volterra sweep needs.

Given $\lambda^{(m)}$, we form $L = (N-1)\lambda^{(m)}$, solve Eq.~(\ref{eq:volterra}) once per hypothesis at effective drifts $\mu_1-L$ and $\mu_0-L$, form the hazards, and take
\begin{align}
  \widehat\lambda(t) = \max\!\bigl[h_1(t) - h_0(t),\;0\bigr],
  \label{eq:fp_map}
\end{align}
the rectification enforcing that silence is never positive evidence for a threat. The next iterate under-relaxes, $\lambda^{(m+1)} = \omega\widehat\lambda + (1-\omega)\lambda^{(m)}$, with $\omega = \min(0.5,\,1.5/N)$ for the first two iterations and $\omega = 0.5$ thereafter, damping the opening steps at large $N$ where the coupling is stiffest and the warm start is furthest from the fixed point. Iteration stops at $\max_t|\lambda^{(m+1)}-\lambda^{(m)}|/\max_t|\lambda^{(m+1)}| < 10^{-4}$, seeded from the uncoupled hazard gap and warm-started at each group size from the converged correction at the previous $N$. The horizon $T_{\max} = \max(8\theta_N,\,3\theta_N^2/\log N)$ resolves the $H=0$ tail that fixes saturation, its second term tracking the order-statistic detection scale of Eq.~(\ref{eq:group_T1}). The dyad converges in twelve iterations and $N=100$ in twenty-five.

Two safeguards are needed at late times. The hazard $h_H = f_H/S_H$ becomes unresolvable once the $H{=}1$ survivor pool is small, since numerator and denominator are then both at quadrature noise level, so beyond that point $\lambda$ is held at its last resolved value. The tail is correspondingly read past the early transient and inside the resolved region, and by extrapolation rather than as a window mean, since $L^{(N)}(t)$ is still decaying there and a mean sits above the limit it is approaching. At $N=2$ the window mean is $0.215$ against an extrapolated $0.164$, a $30\%$ difference on a quantity whose closed form is $0.172$.

\begin{figure}[t!]
  \centering
  \includegraphics[width=0.65\linewidth]{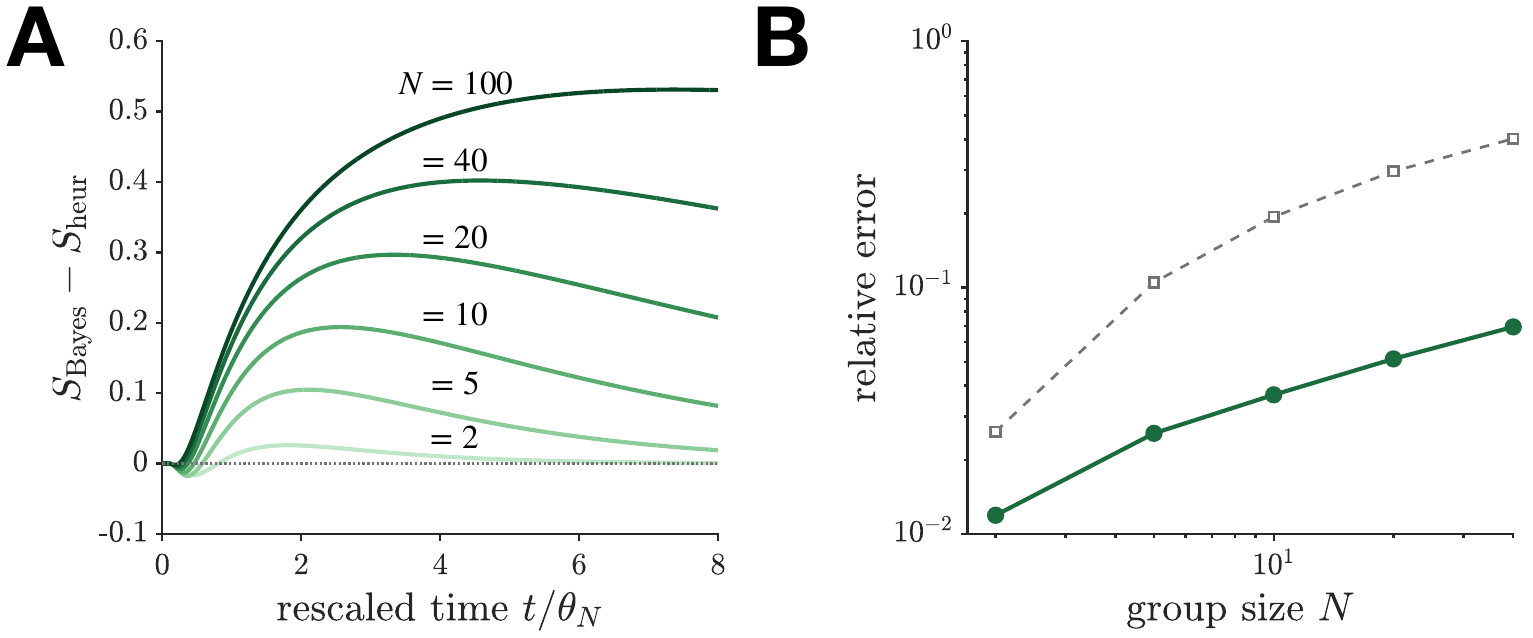}
  \caption{\textbf{Quantifying the departure and cost of the constant-$\alpha$ heuristic against the full Bayesian fixed point.} The heuristic replaces the time-varying correction $\lambda^{(N)}(t)$ by the constant $\alpha = L_\infty(N)$ of Eq.~(\ref{eq:Linf_supp}), both evaluated at the same threshold with the same Durbin primitive, so the comparison isolates the constant-versus-time-varying correction alone. (\textbf{A})~Pointwise discrepancy $S_{\mathrm{Bayes}}(t) - S_{\mathrm{heur}}(t)$ for $H{=}1$ against rescaled time, for $N=2,5,10,20,40,100$ (light to dark). The gap opens through the early transient, where the constant is imposed while the true correction is still rising, and its peak moves later with group size, from $t/\theta_N = 1.8$ at $N=2$ to $7.4$ at $N=100$. It closes within the decision window at small $N$, but from $N=20$ on has not closed by the edge of the plot, which is why the supremum norm in panel B grows. The brief negative excursion near $t/\theta_N \approx 0.4$ is the heuristic over-discounting before the rising correction overtakes the constant. When $H{=}0$ the discrepancy stays at the order of $10^{-3}$ at every $N$ and is omitted. (\textbf{B})~Error in the quantity the heuristic is used for, the group detection time $\bar T_{(1)} = \int S(t\mid H{=}1)^N\,\dd t$ (green circles), growing from $1.2\%$ at $N=2$ to $9.6\%$ at $N=100$, comparable to the saddle-point approximation of Eq.~(\ref{eq:group_T1}) and, like it, never the source of a reported number. The maximum pointwise gap (grey squares) reaches $53\%$ at $N=100$, but a supremum norm on a steep monotone curve is dominated by a small shift in time through the transient, and no result in the paper reads $S(t)$ pointwise.}
  \label{fig:S2}
\end{figure}

Replacing the converged $\lambda^{(N)}(t)$ by the constant $\alpha = L_\infty(N)$ of Section~\ref{sec:Linf_supp} is the defining approximation of the heuristic model, and Fig.~\ref{fig:S2} measures what it costs. The two are not pointwise close and are not expected to be, since the constant is imposed from $t=0$ while the true correction first rises through the early-time transient of Section~\ref{sec:hump}. What matters is the error left in the quantity the heuristic computes, the group detection time $\bar T_{(1)} = \int S(t\mid H{=}1)^N\,\dd t$, and that error grows with group size to $9.6\%$ at $N=100$, the same order as the saddle-point approximation of Eq.~(\ref{eq:group_T1}) and, like it, never the source of a quoted number. When $H=0$ the two agree to the order of $10^{-3}$ at every group size, since the survivor is already drifting away from the barrier and crossing at a vanishing rate, so an extra downward drift of either form has almost nothing left to suppress.

\section{Closed form for the saturation ceiling}
\label{sec:Linf_supp}

\begin{figure}[t!]
  \centering
  \includegraphics[width=0.65\linewidth]{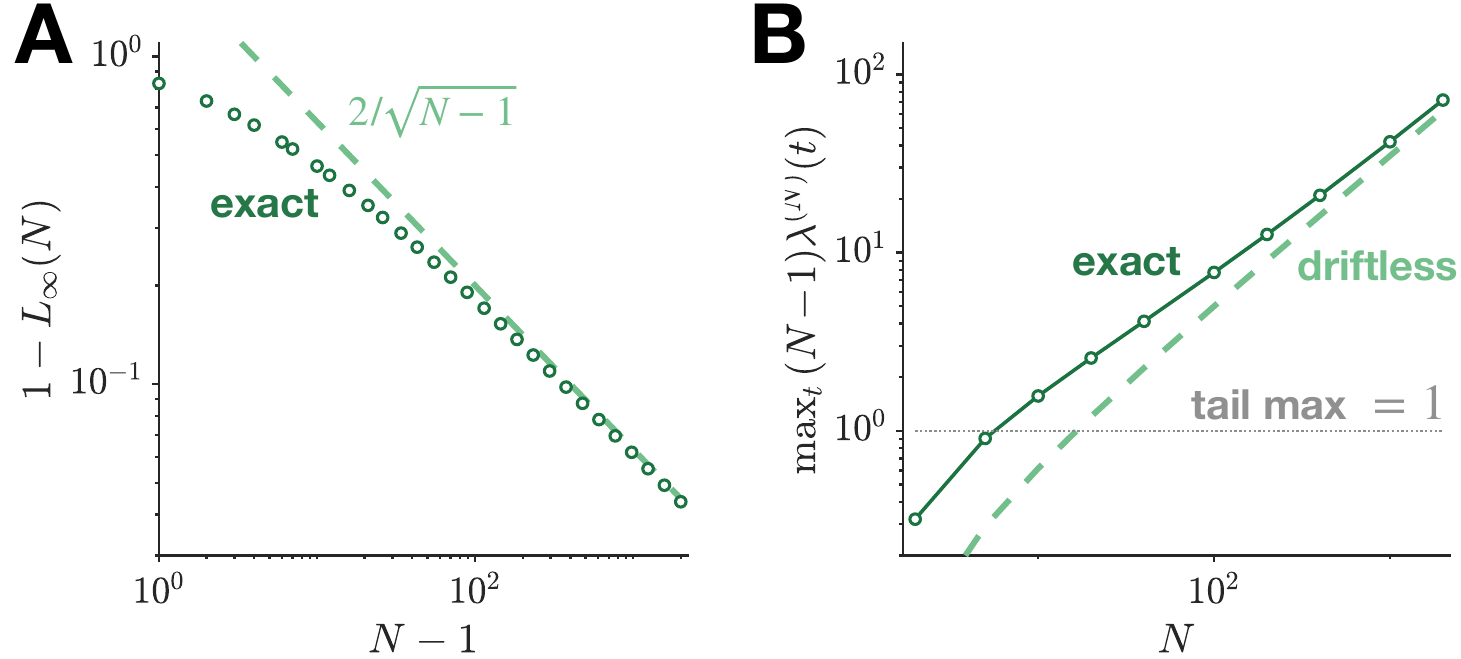}
  \caption{\textbf{Closed-form validation of the saturation ceiling and the hump scaling.} (\textbf{A})~Approach of the ceiling to unity, $1 - L_\infty(N)$ against $N-1$ on logarithmic axes. Circles are the closed form of Eq.~(\ref{eq:Linf_supp}); the dashed line is the leading asymptote $2/\sqrt{N-1}$, of slope $-\tfrac12$. The markers approach it from below, the offset at small $N$ being the $+2/k$ correction. (\textbf{B})~Divergence of the early-time hump, the peak total survival drift against $N$. Circles and solid line are the matched-asymptotics form of Eq.~(\ref{eq:hump_exact}); the dashed line is the driftless leading term $(N-1)C_0/\theta_N^2$ with $C_0 = 1.0207$, approached from above as $\beta_N \to 0$, the ratio falling from $4.1$ at $N=2$ to $1.1$ at $N=2000$. The dotted line marks the tail ceiling of panel A. Unlike the tail, the hump grows without bound.}
  \label{fig:S1}
\end{figure}

The total survival drift $L^{(N)}(t) = (N-1)\lambda^{(N)}(t)$ approaches a finite late-time value. For a constant-drift diffusion with drift $\nu$ and variance rate $2$ against a fixed barrier, the first-passage hazard has large-time limit $\nu^2/4$ when $\nu>0$ and $0$ when $\nu\le0$: a survivor pushed toward threshold crosses at a constant asymptotic rate, one pushed away is asymptotically trapped. At late times the effective drifts are $\tilde\mu_1 = 1-L_\infty$ and $\tilde\mu_0 = -1-L_\infty$. Since $\tilde\mu_0<0$ always, the $H=0$ hazard vanishes and the entire late-time correction is carried by the $H=1$ survivor hazard, so the closure reduces to
\begin{align}
  L_\infty = (N-1)\,\frac{\tilde\mu_1^2}{4} = \frac{(N-1)(1-L_\infty)^2}{4},
  \label{eq:Linf_quadratic}
\end{align}
the quadratic $k(1-L_\infty)^2 = 4L_\infty$ with $k=N-1$, whose roots are $1 + 2/k \pm 2\sqrt{k+1}/k$. The physical root lies in $(0,1)$; the other exceeds unity and forces $\tilde\mu_1<0$. Hence
\begin{align}
  L_\infty(N) = \frac{(k+2) - 2\sqrt{k+1}}{k}, \qquad k = N-1.
  \label{eq:Linf_supp}
\end{align}
At $N=2$ this is $3 - 2\sqrt2 \approx 0.1716$, and the sequence $0.172,\,0.382,\,0.520,\,0.635,\,0.727,\,0.818$ at $N=2,5,10,20,40,100$ matches the extrapolated Durbin tails. For large $k$,
\begin{align}
  L_\infty(N) = 1 - \frac{2}{\sqrt{k}} + \frac{2}{k} + O\!\bigl(k^{-3/2}\bigr),
  \label{eq:Linf_asympt_supp}
\end{align}
so the ceiling reaches unity only as $N\to\infty$, and slowly: $1-L_\infty$ against $N-1$ on logarithmic axes has slope $-\tfrac12$, with the $+2/k$ correction producing the mild upward curvature at small $N$ (Fig.~\ref{fig:S1}A).

\section{The early-time hump: matched asymptotics}
\label{sec:hump}

The total survival drift is not monotone. It settles to $L_\infty(N)<1$, but first rises through an early-time hump that, unlike the tail, is not bounded by unity and diverges with $N$. We establish that the hump is a distinct object with a different scaling, so its overshoot is not mistaken for a failure of saturation.

\subsection{Reduction to the $H=1$ hazard peak}

The hump sits in a boundary layer near the $H=1$ first-passage mode. There the $H=0$ survivor drifts away from the barrier, so $h_0\ll h_1$ and the closure reduces to $\lambda^{(N)}(t)\approx h_1(t)$. The hump height is therefore $(N-1)$ times the peak single-agent hazard at effective drift $\tilde\mu_1 = 1 - L_\infty(N)$. We isolate that peak in two steps: the zero-drift hazard fixes a universal constant and the $\theta^{-2}$ scaling, and a finite-drift correction collapses onto a one-parameter family in $\beta=\tilde\mu_1\theta$.

\subsection{The zero-drift boundary-layer constant}

Let $\mathring h$ be the hazard of a driftless diffusion to level $\theta$, with $\mathring f(t) = \theta(4\pi t^3)^{-1/2}e^{-\theta^2/4t}$ and $\mathring S(t) = \mathrm{erf}(\theta/\sqrt{4t})$. The substitution $u = \theta^2/(4t)$ collapses $\theta^2\mathring h$ onto a function of $u$ alone,
\begin{align}
  \theta^2 \mathring h(t) = \Phi(u), \qquad \Phi(u) = \frac{(4u)^{3/2}\,e^{-u}}{\sqrt{4\pi}\,\mathrm{erf}(\sqrt u)},
  \label{eq:Phi_u}
\end{align}
proving the zero-drift peak hazard scales exactly as $\theta^{-2}$. The maximum is at the root of
\begin{align}
  \frac{3}{2u} - 1 - \frac{1}{\sqrt\pi}\,\frac{e^{-u}/\sqrt u}{\mathrm{erf}(\sqrt u)} = 0,
  \label{eq:ustar}
\end{align}
namely $u^\star = 1.30437$, giving $C_0 \equiv \max_t \theta^2\mathring h(t) = 1.0207$ at $t^\star = 0.1917\,\theta^2$. The peak hazard is thus $C_0/\theta^2$, not the $O(1/\theta)$ the density prefactor might suggest.

\subsection{Finite-drift correction and hump scaling}

At drift $\nu$ the density acquires the factor $\exp(\nu\theta/2 - \nu^2 t/4)$, whose exponent in $u$ is $\beta/2 - \beta^2/(16u)$ with $\beta \equiv \nu\theta$. The peak hazard therefore collapses onto a universal function of $\beta$,
\begin{align}
  \max_t \theta^2 h_\nu(t) = G(\beta), \qquad G(0) = C_0,
  \label{eq:Gbeta}
\end{align}
verified numerically to be $\theta$-independent to five digits, with $G(\beta)/C_0 = 1 + 0.558\beta + 0.136\beta^2 + O(\beta^3)$. The hump height is then exactly
\begin{align}
  \max_t (N-1)\lambda^{(N)}(t) = \frac{(N-1)\,G(\beta_N)}{\theta_N^2}, \qquad \beta_N = \bigl(1 - L_\infty(N)\bigr)\theta_N,
  \label{eq:hump_exact}
\end{align}
reproducing the converged Durbin peak to all digits. The drift strength vanishes at the ceiling, $\beta_N \sim (\theta_1 + \log N)/\sqrt{N-1} \to 0$, so $G(\beta_N)\to C_0$ and
\begin{align}
  \max_t (N-1)\lambda^{(N)}(t) \sim \frac{4 C_0\,(N-1)}{(\theta_1 + \log N)^2} \sim \frac{4 C_0\,N}{(\log N)^2}.
  \label{eq:hump_scaling}
\end{align}
The hump diverges as $N/(\log N)^2$ while the tail saturates (Fig.~\ref{fig:S1}B). The divergence does not undermine the constant-$\alpha$ approximation. The hump is a brief burst of social evidence at the instant a typical $H=1$ cohort crosses, acting over a window that narrows as the peak grows, and it controls no steady-state quantity. Every saturation statement in the paper refers to the tail value $L_\infty(N)$, the quantity identified with $\alpha$, and never to the peak.

\section{Detection-time speedup and threshold scaling}
\label{sec:largeN_supp}

\subsection{Extreme-value detection time}

The group first-departure time has mean $\bar T_{(1)} = \int_0^\infty S(t\mid H{=}1)^N\,\dd t$ in response to a real threat. For large $N$ the integrand is dominated by small $t$, where the inverse-Gaussian cumulative has the short-time form $\PP(T<t\mid H{=}1) \sim \exp[-(\theta-\tilde\mu_1 t)^2/2\sigma^2 t]$. Expanding the exponent,
\begin{align}
  \frac{(\theta-\tilde\mu_1 t)^2}{2\sigma^2 t}
  = \frac{\theta^2}{2\sigma^2 t} - \frac{\tilde\mu_1\theta}{\sigma^2} + \frac{\tilde\mu_1^2 t}{2\sigma^2},
  \label{eq:T1_exponent_supp}
\end{align}
so the integral concentrates where the double exponential turns over, at $\theta^2/(2\sigma^2 t_\star) = \log N + \tilde\mu_1\theta/\sigma^2 + O(\log\log N)$, giving
\begin{align}
  \bar T_{(1)} \sim \frac{\theta^2}{2\,\sigma^2\,\log N},
  \label{eq:T1_supp}
\end{align}
which is Eq.~(\ref{eq:group_T1}). Pooling $N$ independent streams shrinks the detection time as $1/\log N$. The discounting rate does not appear at leading order. Of the three terms in Eq.~(\ref{eq:T1_exponent_supp}), only the first carries $\theta^2/t$ and so grows like $\log N$ at the saddle point, and it is free of $\tilde\mu_1$. The second is constant in $t$ and enters as the additive shift $\tilde\mu_1\theta/\sigma^2$ inside the logarithm, a correction of relative size $1/\log N$. The third is proportional to $t_\star$ and therefore smaller by a further factor of $\log N$. At fixed threshold the leading-order detection time is thus independent of $\alpha$, which survives only in the subleading shift. The dominant dependence runs instead through $\theta(\alpha)$, since $\theta$ enters squared while the threshold scaling of Eq.~(\ref{eq:threshold_supp}) makes it fall as $1/(1+\alpha)$, so at matched false alarm rate a strong discounter detects faster by roughly $(1+\alpha)^2$.

\subsection{Threshold scaling and branching-ratio pinning}

The group false alarm rate is $q_N = 1-(1-q_\alpha(\theta))^N \approx N q_\alpha(\theta)$ for small $q_\alpha = e^{-(1+\alpha)\theta}$. Holding it at the solitary value requires $N e^{-(1+\alpha)\theta_N} = e^{-\theta_1}$, that is
\begin{align}
  \theta_N = \frac{\theta_1 + \log N}{1+\alpha},
  \label{eq:threshold_supp}
\end{align}
and when each agent attends a fixed neighborhood of size $K$, $N$ is replaced by $K$. At this threshold the false alarm branching ratio is
\begin{align}
  b_{H=0} = K q_\alpha(\theta_K) = K e^{-\theta_1} K^{-1} = e^{-\theta_1} = q_1,
  \label{eq:bH0_supp}
\end{align}
independent of group size. Under a real threat every surviving neighbor sits near threshold and the kick clears all $K$ of them, giving $b_{H=1} = K$, so the discriminability $b_{H=1}/b_{H=0} = K/q_1$ grows linearly with the attended count. Anchoring on the $K=1$ case, $b_{H=0}$ carries no dependence on $\alpha$, since the fixed level is the true solitary false alarm rate. The benefit of discounting is therefore not a larger discriminability at fixed threshold but a lower threshold at matched false alarm rate, and hence faster detection.

\section{Empirical identification: data and methods}
\label{sec:identification_supp}

Section~\ref{ssec:datasource} describes the data. Section~\ref{ssec:rates} extracts the two response rates and states what they pool over. Section~\ref{ssec:estimator} gives the estimator of $\alpha$ and its intervals. Section~\ref{ssec:poolcount} identifies the pooling count from the latency shape, which the previous two depend on, and checks the assumptions behind it. Section~\ref{ssec:area} turns to the scaling with shoal size, the one prediction that plays no part in the fit.

\subsection{Data source and provenance}
\label{ssec:datasource}

We use the publicly archived field dataset of Pacher et al.~\cite{pacher2025better}, recording collective escape responses of wild sulphur molly (\textit{Poecilia sulphuraria}) shoals to natural aerial disturbances at a sulphidic spring system in southern Mexico. Their Data~S1 is a per-event table in which each row is an overhead event scored as a predator attack or a harmless flyby. We use the event class, the recording identifier, the bout index, the shoal area, the true-positive flag, and the latency to the first repeat wave (recorded in video frames at $25$~fps). We take no preprocessing step beyond unit conversion and the exclusion of events with a missing latency from the timing analysis. The sample comprises $n=177$ attacks ($127$ eliciting a response, $50$ not) and $n=81$ flybys ($26$ eliciting a response), with $n=125$ attacks carrying a valid latency, drawn from $18$ recordings. Twenty-three flybys also carry a latency; these are false-alarm response times rather than detections and are excluded from the timing analysis, which uses the $125$ attack latencies alone.

\begin{figure}[t!]
  \centering
  \includegraphics[width=\linewidth]{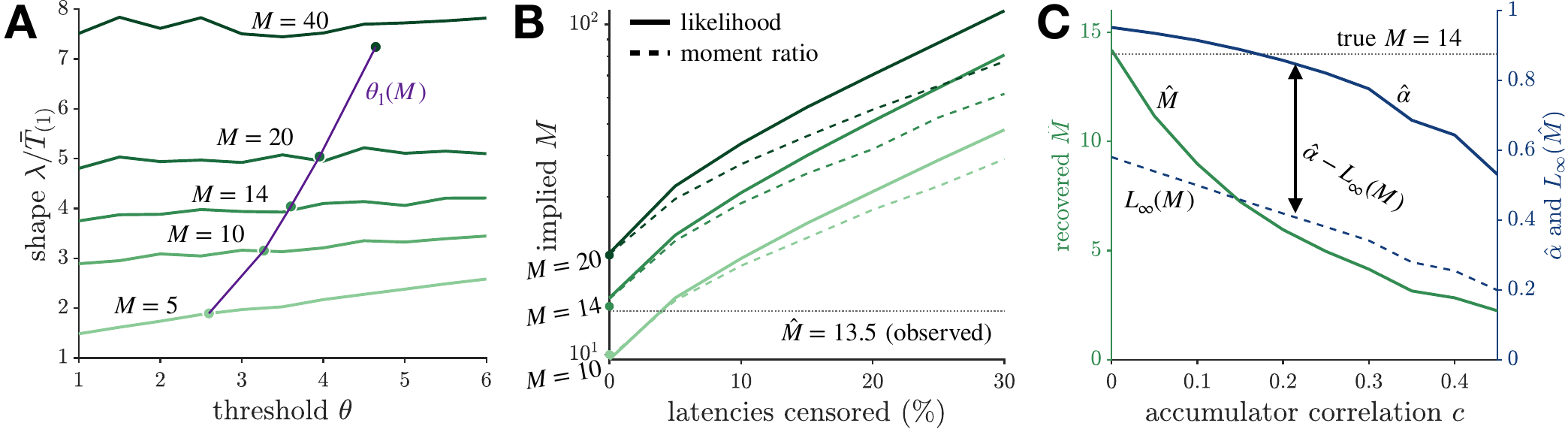}
  \caption{\textbf{Robustness of the pooling count.} (\textbf{A})~The scale-free statistic $\lambda/\bar T_{(1)}$ against threshold, for pools of $M = 5, 10, 14, 20, 40$ (light to dark), with the self-consistent points $\theta = \theta_1(M)$ marked and joined (purple). The statistic is not threshold-free at small pools, with relative spread across $\theta\in[1,6]$ of $54\%$, $18\%$, $12\%$, $8\%$, and $5\%$. It need not be calibrated at an arbitrary threshold, since the false alarm rate fixes $\theta_1(M) = -\log[1-(1-q)^{1/M}]$, so the calibration runs along the purple locus and closes on itself. (\textbf{B})~Sensitivity to right-censoring. Latencies are simulated at a known pooling count ($M = 10, 14, 20$) and at the self-consistent threshold for that count, the upper tail censored, and the implied $\hat M$ read back by the moment ratio (dashed) and by a scale-free likelihood cross-check (solid). Because simulation and calibration share that threshold, the uncensored estimate returns the true count to within a few percent, so the curves begin at their markers and the displacement across the panel is attributable to censoring. Truncating the right tail inflates the shape statistic and so inflates the estimate, and at $5\%$ censoring a true $M=14$ reads as $22$ by the moment ratio, at $20\%$ as $42$, with the likelihood cross-check agreeing at $5\%$ and reading higher at $20\%$. The molly latencies show no censoring signature, so the relevant point is the left edge; the curves bound how far an undetected censoring could move the estimate, not how far it does. (\textbf{C})~Correlated accumulators, the assumption behind the calibration. Equicorrelated streams reduce the number of effectively independent detectors, so the recovered $\hat M$ falls with the correlation $c$ (green, left axis), halving by $c = 0.2$, and $\hat\alpha$ falls with it (blue solid, right axis). The individually Bayesian benchmark $L_\infty(\hat M)$ falls too (blue dashed), but the gap between them does not narrow monotonically. As correlation drives $\hat M$ down the benchmark falls faster at first, so the gap widens before turning over near $c = 0.25$, and across the range in which $\hat M$ stays within the calibrated grid ($c \le 0.45$) it varies only between $0.33$ and $0.44$. The point estimate of $\hat M$ is therefore conditional on independence, while the sign of the over-discounting result is insensitive to it.}
  \label{fig:S3}
\end{figure}

\begin{figure}[t!]
  \centering
  \includegraphics[width=\linewidth]{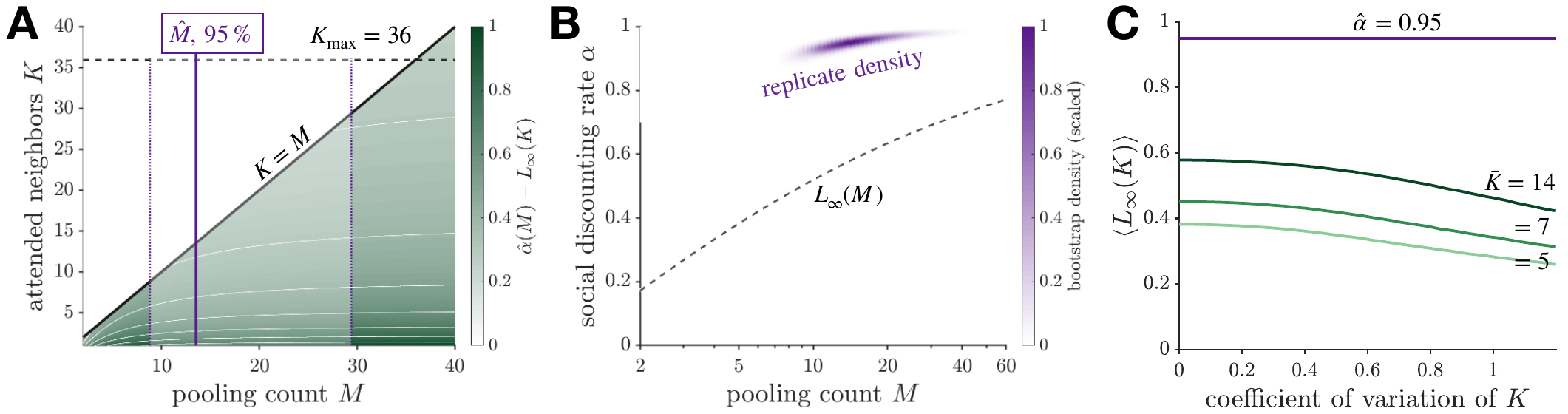}
  \caption{\textbf{The over-discounting conclusion does not require the attended count to be identified.} (\textbf{A})~The gap $\hat\alpha(M) - L_\infty(K)$ over the admissible wedge $K \le M$, white contours at $0.1$ intervals. Black diagonal, $K = M$; dashed, the admissibility ceiling $K_{\max} = 36$; purple line and shaded band, $\hat M$ and its $95\%$ interval. The gap is positive over the entire wedge, with a minimum of $0.26$ on the diagonal where the attended count is largest, so no choice of attended count within the pooling count reverses the conclusion. The ceiling does not bind over the range plotted, since $K \le M \le 40$ throughout. (\textbf{B})~Joint bootstrap density over $(M,\alpha)$, $B = 4000$ recording-by-bout clusters resampled and stratified by event type, with $L_\infty(M)$ drawn across it (dashed). Since $L_\infty$ is increasing and $K \le M$, that curve is the largest value the benchmark can take at each pooling count; every replicate lies above it, $P(\hat\alpha > L_\infty) = 1.000$, with a minimum excess of $0.20$. The cloud is right-skewed in $M$, its median $14.3$ against the point estimate $13.5$. (\textbf{C})~Heterogeneity in the attended count. Because $L_\infty$ is concave over this range, $\E[L_\infty(K)] < L_\infty(\E[K])$, so dispersion in $K$ lowers the benchmark and widens the gap: at $\bar K = 7$ it falls from $0.451$ at zero dispersion to $0.405$ and $0.342$ at coefficients of variation $0.6$ and $1.0$. Curves are $\bar K = 5, 7, 14$ (light to dark); the horizontal line is $\hat\alpha = 0.95$.}
  \label{fig:S4}
\end{figure}

\begin{figure}[t!]
  \centering
  \includegraphics[width=\linewidth]{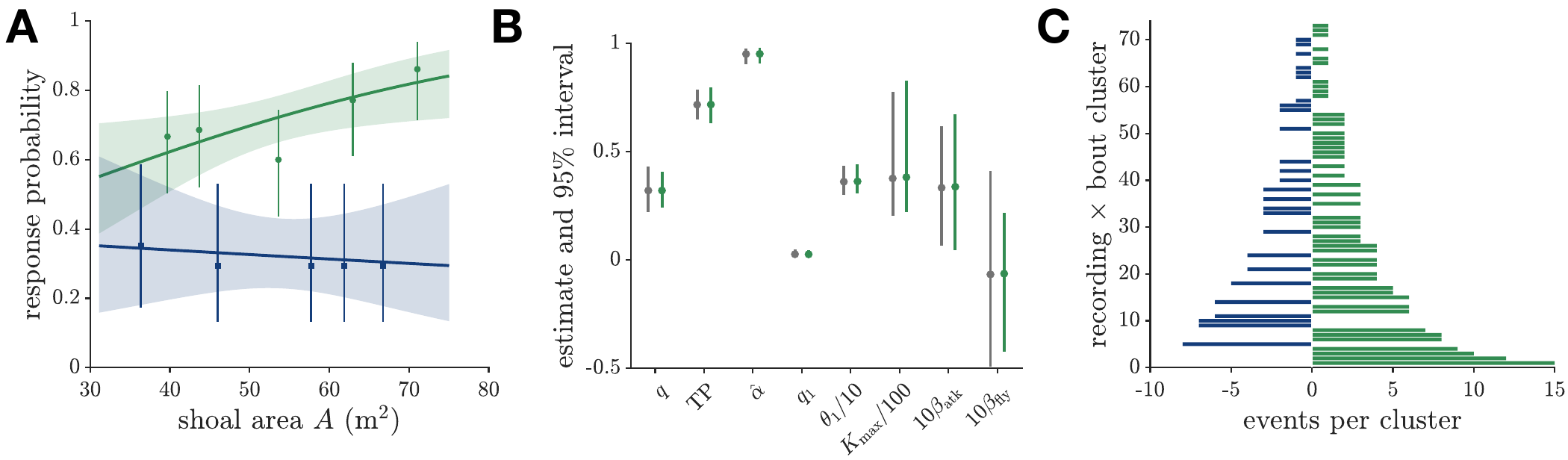}
  \caption{\textbf{Rate estimation and what the intervals rest on.} (\textbf{A})~Response probability against shoal area, attacks (green circles) and flybys (blue squares), as quintile means with Wilson $95\%$ intervals sized by bin count, with logistic fits and $95\%$ ribbons. The flyby slope is flat ($-0.006$, $P = 0.78$) and the attack slope rises ($+0.033$, $P = 0.019$), but a pooled model with an event-type interaction does not resolve them apart ($z = 1.54$, $P = 0.12$). (\textbf{B})~Bootstrap $95\%$ intervals for every rate-derived quantity and for the two area slopes, resampling events independently (grey) and resampling recording-by-bout clusters stratified by event type (green); $\theta_1$, $K_{\max}$, and the slopes are rescaled to share the axis. The clustered intervals are the ones reported throughout. They are not uniformly wider than the independent ones, running from $24\%$ wider for $\mathrm{TP}$ to $21\%$ narrower for $q$, so the independent interval is not a conservative stand-in for either. Neither slope changes sign upon clustering: the attack slope stays strictly positive and the flyby slope still spans zero. (\textbf{C})~Cluster structure, events per cluster with attacks to the right and flybys to the left, over the $73$ clusters. Only $26$ carry flybys, so the effective sample behind $q$ is far smaller than the raw count of $81$ suggests, which is why the clustered resample and not the event-level one is the right basis for every $q$-derived interval.}
  \label{fig:S5}
\end{figure}

\begin{figure}[t!]
  \centering
  \includegraphics[width=\linewidth]{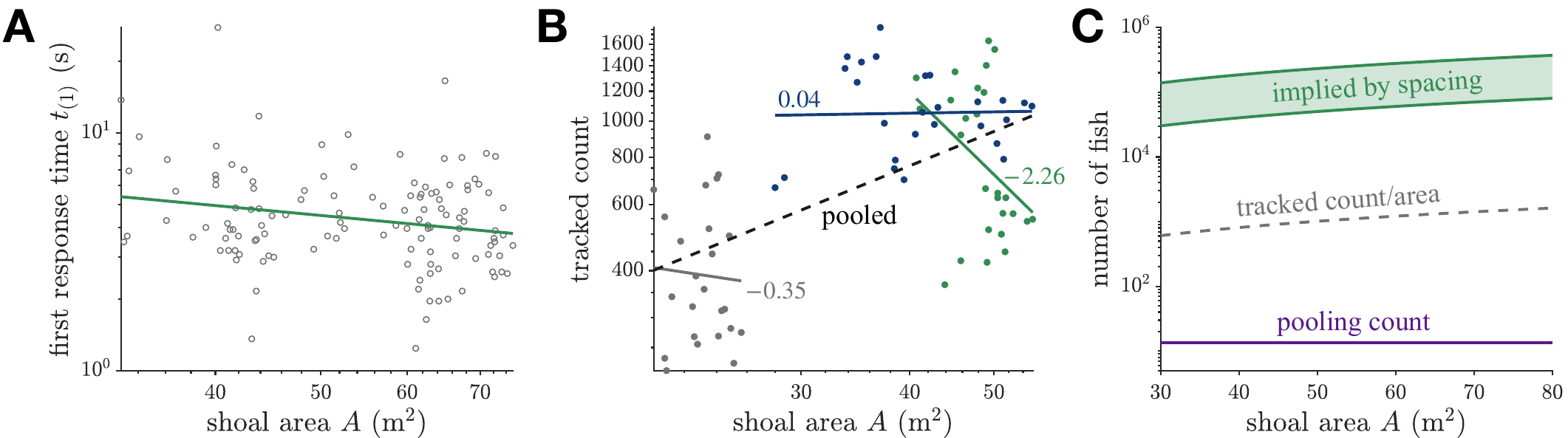}
  \caption{\textbf{Shoal area, headcount, and shoal density.} (\textbf{A})~First response time against shoal area on logarithmic axes, with the power-law fit, over the $125$ timed attacks. The slope is $-0.43$, negative as a first-passage minimum requires, but its recording-by-bout clustered $95\%$ interval crosses zero, so the panel is a sign check and does not bound the pooling count. Shoal area is nearly constant within a recording-bout (within-cluster SD of $\log A$ is $0.018$ against a total of $0.234$, under $1\%$ of the variance), so the slope is identified across the $42$ recordings that carry a timed attack rather than across the $125$ events, and the width of the interval reflects that rather than any deficiency in the timing data. (\textbf{B})~Tracked count against area in the companion structural dataset, by site (colors) with the pooled fit (dashed). The pooled slope of $0.94$ reads as constant density but is entirely between-site; the within-site slopes are $-2.26$, $+0.04$, and $-0.35$, so shoal area measures spread rather than headcount. (\textbf{C})~Headcount implied by the measured nearest-neighbor distance, median $15.8$~mm, between the Poisson and close-packed readings (green band), against the fitted pooling count (purple). The tracked count divided by the recorded area (grey dashed) lies some fiftyfold below the spacing-implied density and is inconsistent with that spacing, so it is a subset rather than a census. The pooling count is about $10^{-4}$ of the fish present, the quantitative form of the correlated and occluded visual fields argument.}
  \label{fig:S6}
\end{figure}

\subsection{Response rates and per-unit crossing probabilities}
\label{ssec:rates}

The two group-level observables are the true-positive rate $\mathrm{TP} = 127/177 = 0.718$ and the false-alarm rate $q = 26/81 = 0.321$ (Wilson $95\%$ interval $[0.229,0.429]$, unclustered). A repeat wave is recorded when any fish in the observed shoal responds, so both rates pool over the $M$ responders whose earliest crossing registers as a group response, not over the $K$ neighbors any one fish attends. Inverting $1-(\text{group miss}) = (1-\text{per-unit response})^M$ recovers
\begin{align}
  q_{\rm ind} = 1-(1-q)^{1/M}, \qquad p_{\rm ind} = 1-(1-\mathrm{TP})^{1/M}, \qquad \mathrm{miss}_{\rm ind} = 1-p_{\rm ind},
  \label{eq:pool_supp}
\end{align}
where $M$ is estimated in Section~\ref{ssec:poolcount} and is not fixed by hand. The nonzero per-unit miss is supplied by the finite observation window.

\subsection{Estimator of the discounting rate}
\label{ssec:estimator}

In the heuristic model the per-unit false-alarm and miss probabilities are $q_{\rm ind} = e^{-2(1+\alpha)\theta/\sigma^2}$ and $\mathrm{miss}_{\rm ind} = e^{-2(1-\alpha)\theta/\sigma^2}$. Their log-ratio cancels the common factor $2\theta/\sigma^2$,
\begin{align}
  r \equiv \frac{\log q_{\rm ind}}{\log \mathrm{miss}_{\rm ind}} = \frac{1+\alpha}{1-\alpha}
  \qquad\Longrightarrow\qquad
  \hat\alpha = \frac{r-1}{r+1},
  \label{eq:sigmafree_supp}
\end{align}
so neither the threshold nor the noise scale enters. At $\hat M = 13.5$ this gives $\hat\alpha = 0.95$, with per-unit false alarm rate $q_1 = 0.028$, threshold $\theta_1 = -\log q_1 = 3.57$, and admissibility ceiling $K_{\max} = \log(1-q_1)/\log(1-q_1^2) = 36$.

Confidence intervals are joint bootstrap percentile intervals over $B = 4000$ replicates, propagating rate sampling error and the uncertainty in $\hat M$ together, since $\hat\alpha$ depends on $M$. The resampling is over clusters rather than events, keyed on the recording identifier combined with \texttt{bout\_id}, since the bout index restarts within each recording, and stratified by event type, since an unstratified draw can empty the flyby stratum and return undefined rates. That gives $73$ clusters, $47$ carrying attacks and only $26$ carrying flybys, so the effective sample behind $q$ is far smaller than the raw count of $81$ suggests. The independent-events interval is not a conservative stand-in for the clustered one either, the clustered intervals running from $24\%$ wider for $\mathrm{TP}$ to $21\%$ narrower for $q$ (Fig.~\ref{fig:S5}B).

The comparison against the individually Bayesian benchmark does not require the attended count to be identified. Since $L_\infty$ is increasing and $K \le M$ by counting, the benchmark at a given pooling count is bounded above by $L_\infty(M)$, and $\hat\alpha$ exceeds that bound over the whole admissible wedge. Three excesses are worth distinguishing, all positive and none interchangeable: $0.38$ at the point estimate $(\hat M, \hat\alpha)$; $0.26$ minimized over the wedge $K\le M$, attained on the diagonal where the attended count is largest; and $0.20$ minimized over the bootstrap replicates, which reach pooling counts well above $\hat M$ and where the excess is smaller because it decreases in $M$. Dispersion in $K$ across individuals lowers the benchmark further, since $L_\infty$ is concave over this range (Fig.~\ref{fig:S4}).

\subsection{Identifying the pooling count from the latency shape}
\label{ssec:poolcount}

The observed latency is the first repeat wave in the shoal, the minimum over $M$ first passages rather than any one unit's single passage. Both the inverse-Gaussian shape parameter $\lambda$ and the mean carry units of time, so $\lambda$ alone says nothing about sharpness and the scale-free statistic is $\lambda/\bar T_{(1)}$. We calibrate $M \mapsto \lambda/\bar T_{(1)}$ by Monte Carlo on a grid $M = 2,\dots,90$ with $2\times10^6$ draws per point and invert by interpolation. The observed value $19.32/4.92 = 3.93$ returns $\hat M = 13.5$.

The threshold is not a free parameter of that calibration. The statistic depends on it, mildly at large $M$ and appreciably at small, with spread across $\theta\in[1,6]$ of $12\%$ at $M=14$ and $54\%$ at $M=5$ (Fig.~\ref{fig:S3}A), so calibrating at an arbitrary threshold moves the estimate. It need not be arbitrary: once $M$ is chosen the false alarm rate fixes the threshold through $\theta_1(M) = -\log[1-(1-q)^{1/M}]$. Running the calibration along that locus closes the loop, and the estimate is then stable to $\pm0.07$ across Monte Carlo seeds. The calibration itself is held at the observed rates across bootstrap replicates, so the interval on $\hat M$ reflects sampling in the latencies and in the two rates rather than uncertainty in the calibration curve.

Identifying the decision time with a single isolated accumulator is unrealistic. At the fitted discounting rate the effective drift is $1-\hat\alpha = 0.05$, so reproducing the observed shape from one accumulator would require $\theta \approx 154$, more than forty times the threshold implied by the rates. Nor does the shape parameter yield a noise scale, since the model carries no timescale of its own and we claim no separate estimate of the signal-to-noise ratio. Only scale-free features of the latencies can identify $M$. A likelihood written on the raw latencies is not scale-invariant, and profiling a free time scale out of it restores invariance but yields no advantage over the ratio, with $95\%$ recovery at $n=125$ of $[10.1,22.6]$ against $[10.3,22.5]$.

Figure~\ref{fig:S3} checks two assumptions behind $\hat M$. The calibration treats the $M$ accumulators as independent, and correlated streams reduce the effective count, so panel C repeats it under equicorrelated accumulators and finds $\hat M$ halving by a correlation of $0.2$ with $\hat\alpha$ following it down. The individually Bayesian benchmark falls too, and faster at first, so the gap widens before turning over near a correlation of $0.25$ and stays between $0.33$ and $0.44$ across the range in which $\hat M$ remains within the calibrated grid. The point estimate is therefore conditional on independence, while the sign of the result is not. Panel B addresses the second assumption, that the $50$ unanswered attacks are misses rather than censored late responses, bounding how far an undetected censoring could move the estimate. The latencies show no censoring signature, decaying smoothly to a maximum of $27.8$~s with a single event within $2$~s of it, where a fixed observation window would leave a pile-up.

\subsection{Area scaling and shoal density}
\label{ssec:area}

The constancy of the false alarm rate in shoal size is an out-of-sample prediction, since neither the area coordinate nor the binning enters the identification. Binned rates use Wilson score $95\%$ intervals. Fitting a logistic in shoal area separately by event type gives a flyby slope of $-0.006$ (SE $0.021$, $P = 0.78$), consistent with a constant rate and not with the increase the naive model requires, while the attack slope is $+0.033$ (SE $0.014$, $P = 0.019$). These standard errors treat events as independent, and the clustered counterparts in Fig.~\ref{fig:S5}B change neither sign. A pooled model with an event-type interaction does not resolve the two apart ($z = 1.54$, $P = 0.12$), so each slope is supported on its own while the contrast between them is not (Fig.~\ref{fig:S5}A).

The detection-time scaling is a weaker test and we do not present it as an independent one, since the latency distribution supplies $\hat M$ and those data are inside the fit. What remains is the dependence of mean latency on shoal area, which two things limit. Its log-log slope is $-0.43$, negative as a first-passage minimum requires, but the clustered $95\%$ interval crosses zero, so it checks the sign without bounding $M$ (Fig.~\ref{fig:S6}A). More seriously, the companion structural dataset shows that area does not track headcount within a site, since regressing log count on log area pools to a slope of $0.94$ that reads as constant density but is entirely between-site, the within-site slopes being $-2.26$, $+0.04$, and $-0.35$ (Fig.~\ref{fig:S6}B). Shoal area measures how spread out the fish are rather than how many there are, so we do not use it as a proxy for pool size.

The companion dataset gives a median nearest-neighbor distance of $15.8$~mm at a median body length of $13.5$~mm, a spacing of $1.17$ body lengths. Read as a two-dimensional Poisson field that is $1006$ fish per square meter, and at hexagonal close packing $4644$, so a shoal of the median observed area holds between $5.7\times10^4$ and $2.6\times10^5$ fish (Fig.~\ref{fig:S6}C). The fitted pooling count of $13.5$ is therefore about $10^{-4}$ of the fish present, as expected when neighboring fish share correlated and occluded visual fields. The tracked count divided by the recorded area gives only $20$ fish per square meter, some fiftyfold below the spacing-implied density and inconsistent with a nearest-neighbor distance of $15.8$~mm, so that count is a subset rather than a census and we use the spacing.

\end{document}